\documentclass[
prc,%
10pt,%
final,%
notitlepage,%
oneside,%
twocolumn,%
nobibnotes,%
nofootinbib,
superscriptaddress,%
floatfix,%
floatfix,%
showkeys,%
showpacs]%
{revtex4}
\usepackage{color}
\usepackage{amsfonts}
\usepackage{amsbsy}
\usepackage{mathrsfs}
\usepackage{graphicx}
\def\lsim{\mathrel{\rlap{
\lower4pt\hbox{\hskip-3pt$\sim$}}
    \raise1pt\hbox{$<$}}}     
\def\gsim{\mathrel{\rlap{
\lower4pt\hbox{\hskip-3pt$\sim$}}
    \raise1pt\hbox{$>$}}}     
\def\scr#1{\mbox{\scriptsize #1}}
\def\thalf{{\textstyle{\frac{1}{2}}}}
\def\oneqt{{\textstyle{\frac{1}{4}}}}
\begin{document}
\title{	
Global $\Lambda$ polarization in heavy-ion collisions at energies 2.4--7.7 GeV:
\\
Effect of Meson-Field Interaction
} 
\author{Yu. B. Ivanov}\thanks{e-mail: yivanov@theor.jinr.ru}
\affiliation{Bogoliubov Laboratory for Theoretical Physics, 
Joint Institute for Nuclear Research, Dubna 141980, Russia}
\affiliation{National Research Nuclear University "MEPhI", 
Moscow 115409, Russia}
\affiliation{National Research Centre "Kurchatov Institute",  Moscow 123182, Russia} 
\author{A. A. Soldatov}
\affiliation{National Research Nuclear University "MEPhI",
Moscow 115409, Russia}
\begin{abstract}
Based on the three-fluid model, the global $\Lambda$  polarization 
in Au+Au collisions at 2.4 $\leq\sqrt{s_{NN}}\leq$ 7.7 GeV is calculated, including its 
rapidity and centrality dependence.  
Contributions from the thermal vorticity and meson-field term
(proposed by Csernai, Kapusta and Welle) to the global  polarization are considered. 
The results are compared with data from recent and ongoing STAR and HADES experiments.
It is predicted that 
the polarization maximum is reached at $\sqrt{s_{NN}}\approx$ 3 GeV, if the measurements are performed
with the same acceptance. 
The value of the polarization is very sensitive to interplay of the aforementioned 
contributions. In particular, the thermal vorticity results in quite strong increase of the polarization
from the midrapidity to forward/backward rapidities, while 
the meson-field contribution considerably flattens the rapidity dependence. 
The polarization turns out to be very sensitive to details of the equation of state.
While collision dynamics become less equilibrium with decreasing collision energy, 
the present approach to polarization is based on the assumption of thermal equilibrium. 
It is found that equilibrium is achieved at the freeze-out stage, but this equilibration 
takes longer at moderately relativistic energies.
\pacs{25.75.-q,  25.75.Nq,  24.10.Nz}
\keywords{relativistic heavy-ion collisions, 
  hydrodynamics, polarization}
\end{abstract}
\maketitle

\section{Introduction}
\label{Introduction}

Measurements of polarization of particles produced in heavy-ion collisions give us access to a new class of 
collective phenomena, i.e. collective rotation of the nuclear medium. The 
STAR Collaboration  at the Relativistic Heavy Ion Collider
(RHIC) observed nonzero global polarization of $\Lambda$ and $\bar{\Lambda}$ 
at collision energies  7.7 $\leq\sqrt{s_{NN}}\leq$ 200 GeV \cite{STAR:2017ckg,Adam:2018ivw} 
and, recently, multi-strange hyperons \cite{STAR:2020xbm} at 200 GeV. 
Local polarization along the beam direction also was measured \cite{STAR:2019erd}. 
These measurements demonstrated 
rising of the global polarization  with decreasing $\sqrt{s_{NN}}$.

The spin polarization below 7.7 GeV is less explored. 
While a simple extrapolation of this trend suggests
that the global polarization continues to rise as $\sqrt{s_{NN}}$ decreases, we
expect vanishing global polarization at $\sqrt{s_{NN}}=2 m_N$ due to the lack
of system angular momentum. Therefore, a peak in global polarization should 
exist in the region 1.9 $\leq\sqrt{s_{NN}}\leq$ 7.7 GeV. 
Recent model calculations predict this peak
in the different places: at $\sqrt{s_{NN}}\approx$ 3 GeV \cite{Deng:2020ygd,Ivanov:2020udj}  
and  at $\sqrt{s_{NN}}\approx$ 7.7 GeV \cite{Guo:2021udq}.

First data (some of them preliminary) on the global polarization of $\Lambda$ were 
presented in Refs. \cite{STAR:2021beb,Okubo:2021dbt,HADES:SQM2021} for  energies 
3 GeV, 7.2 GeV, and 2.4 GeV, respectively. The first two energy points are obtained within 
STAR fixed-target program (FXT-STAR) at RHIC \cite{Meehan:2017cum}, the third point, by 
HADES Collaboration at GSI Helmholtzzentrum für Schwerionenforschung \cite{HADES:2009aat}. 
These data indicate that the peak in global polarization is reached certainly below the energy of 7.7 GeV.

In this paper we present  calculations of the global $\Lambda$ polarization 
at energies 2.4--7.7 GeV. This energy range covers the energies of the aforementioned FXT-STAR 
and HADES experiments, as well as of the forthcoming experiments at 
the Facility for Antiproton and Ion Research (FAIR) in Darmstadt \cite{Ablyazimov:2017guv} and
Nuclotron-based Ion Collider fAcility (NICA) in Dubna  \cite{Kekelidze:2017ghu}. 

The calculations are performed within the model of the three-fluid dynamics (3FD) \cite{3FD} 
combined with thermodynamic approach to the particle polarization 
\cite{Becattini:2013fla,Becattini:2016gvu,Fang:2016vpj}. 
The simulations are done with three different equations of
state (EoS's): a purely hadronic EoS \cite{Mishustin:1991sp} and two versions
of the EoS with the deconfinement transition \cite{Toneev06},
i.e. a first-order phase transition (1PT) and a crossover
one. The physical input of the present 3FD calculations
is described in Ref. \cite{Ivanov:2013wha}.
A brief report on this study has been already presented in Ref. 
\cite{Ivanov:2020udj}. Here we present results of refined and extended calculations, as 
described in Secs. \ref{polarization in 3FD model} and \ref{Meson-field}. 
The thermodynamic approach based on hadronic degrees 
of freedom \cite{Becattini:2013fla,Becattini:2016gvu,Fang:2016vpj}
well describes the global polarization of hyperons, as  was   
demonstrated by its realizations in various hydrodynamical 
\cite{Karpenko:2016jyx,Xie:2016fjj,Xie:2017upb,Ivanov:2019ern,Ivanov:2019wzg,Ivanov:2020wak,Fu:2020oxj,Sun:2021nsg}
and transport  
\cite{Li:2017slc,Sun:2017xhx,Wei:2018zfb,Shi:2017wpk,Kolomeitsev:2018svb,Vitiuk:2019rfv,Lei:2021mvp} 
models of heavy-ion collisions. 
Though, this thermodynamic approach faces some problems, e.g., in explaining the 
$\Lambda$-$\bar{\Lambda}$ splitting, see recent reviews in Ref. \cite{Becattini:2020ngo,Karpenko:2021wdm}.

\section{Thermalization in nuclear collisions}
\label{Thermalization}

The 3FD model takes into account nonequilibrium at the early stage of nuclear collisions.  
This nonequilibrium stage 
is modeled by means of two counterstreaming baryon-rich fluids (p and t fluids). 
Newly produced particles, dominantly populating the midrapidity region, 
are attributed to a fireball (f) fluid.
These fluids are governed by conventional hydrodynamic equations 
coupled by friction terms in the right-hand sides of the Euler equations.

\begin{figure}[!hbt]
\includegraphics[width=8.5cm]{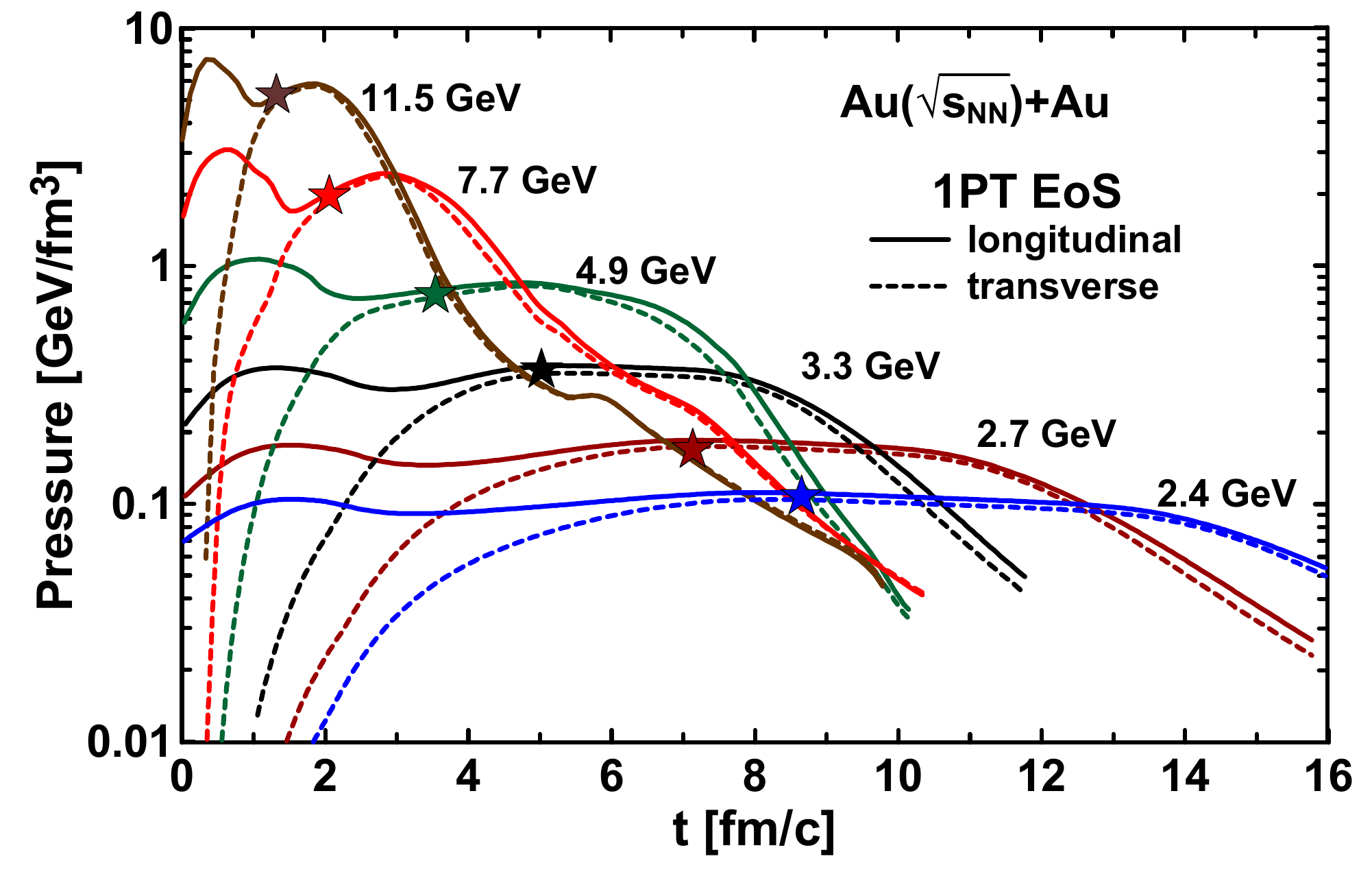}
 \caption{(Color online)
 Time evolution of the longitudinal and transverse pressure 
in the central region of 
Au+Au collision at various 
collision energies ($\sqrt{s_{NN}}$). 
The simulations are performed with the 
1PT EoS. Star symbols on the curves 
mark the time instants of the mechanical equilibration. 
}
\label{fig4m}
\end{figure}

The model \cite{Becattini:2013fla,Becattini:2016gvu,Fang:2016vpj}
used to calculate the global polarization of $\Lambda$ is based on thermodynamic concepts. 
At moderately relativistic energies, 
the thermalization of the matter of colliding nuclei is slow and hence  
the early nonequilibrium stage of nuclear collisions can be quite long. 
Therefore, before proceeding 
to model predictions it is instructive to 
consider degree of the thermalization of the matter at the freeze-out stage. 
Mechanical equilibration in the center 
region of colliding nuclei was studied in Ref. \cite{Ivanov:2019gxm}. 
Criterion of the mechanical equilibration 
is equality of longitudinal  and transverse pressures
with the accuracy no worse than 10\%. 
It is relevant to the nuclear collisions because the leading inequilibrium 
at the initial stage of the collision is associated with anisotropy of the 
momentum distribution along and transverse the beam direction. 
Time evolution of these pressure components 
in the central region of 
Au+Au collision at various collision energies ($\sqrt{s_{NN}}$) is displayed 
in Fig. \ref{fig4m}. The simulations are performed with the 1PT EoS. 
Time instants, when the equilibration 
happens, are marked by star symbols on the curves in Fig. \ref{fig4m}.

A peculiar time evolution of the pressure at the energy of 11.5 GeV (see wiggle $t$ = 5--6 fm/c) 
is a signal of the mixed phase through which the system passes. 
At 7.7 GeV, the mixed phase manifests itself only as a weak irregularity in the evolution, 
since the system quickly passes this phase. 
Results with the crossover EoS are very similar, of course, without these irregularities.  

As seen, the mechanical equilibration is indeed slow at the moderately relativistic energies, 
see Fig. \ref{fig4m}. However, even at  $\sqrt{s_{NN}}=$ 2.42 GeV it is reached 
($\approx$9 fm/c) to the freeze-out stage. The freeze-out stage is extended in time, 
though it is completed at $\approx$20 fm/c from the beginning of the collision. 
The end points of the evolution curves in Fig. \ref{fig4m} correspond to the 
end of the freeze-out stage.  
The mechanical equilibration is of prime importance for applicability of the thermodynamic
model \cite{Becattini:2013fla,Becattini:2016gvu,Fang:2016vpj}.

\begin{figure}[bht]
\includegraphics[width=7.cm]{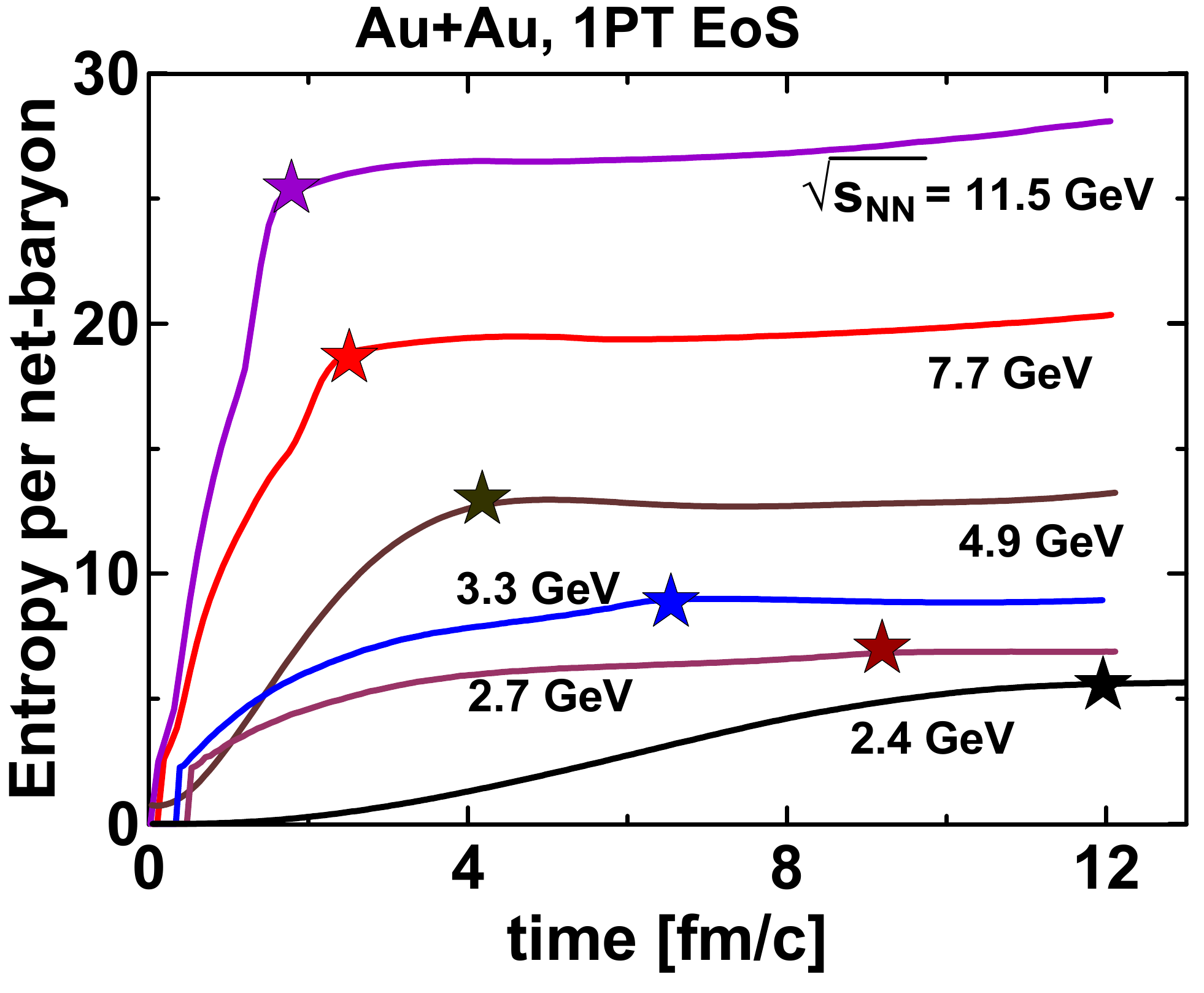}
 \caption{(Color online)
Specific entropy per net baryon ($N_B=2A=$ 394) generated in 
Au+Au collisions  at various energies $\sqrt{s_{NN}}=$ 2.4--11.5 GeV
within  1PT scenario \cite{Toneev06} in the 3FD simulations.
Star symbols on the curves 
mark time instants of the thermalization, i.e. those when 
the rapid growth of entropy is completed.
}
\label{fig4}
\end{figure}

The chemical equilibration and thus thermalization takes longer. Evolution of entropy 
\cite{Ivanov:2016hes}
(Fig. \ref{fig4}) shows that at  $\sqrt{s_{NN}}=$ 2.42 GeV the thermalization takes 
place at $\approx$12 fm/c. Estimation of the 
thermalization within other models 
\cite{Bravina:2008ra,Petersen:2008dd,De:2015hpa,Teslyk:2019ioo,Oliinychenko:2015lva} 
also indicates that it 
takes long time, i.e. of the order of that in the 3FD or even longer, but it is 
completed before the freeze-out stage. 
The success of the statistical model \cite{Andronic:2005yp} at moderate energies also indicates the thermalization at the freeze-out. 

%

\section{Global polarization in 3FD model}
\label{polarization in 3FD model}

In the thermodynamic approach \cite{Becattini:2013fla,Becattini:2016gvu,Fang:2016vpj}, 
particle polarization is related to so-called thermal vorticity defined as 
   \begin{eqnarray}
   \label{therm.vort.}
   \varpi_{\mu\nu} = \frac{1}{2}
   (\partial_{\nu} \beta_{\mu} - \partial_{\mu} \beta_{\nu}), 
   \end{eqnarray}
where $\beta_{\mu}=u_{\nu}/T$, 
$u_{\mu}$ is collective local four-velocity of the matter,  and
$T$ is local temperature.  Here we deal with $u_{\mu}$ and $T$ of the unified fluid 
because the system is equilibrated at the freeze-out stage, as argued in the previous section. 
In the leading order in the thermal vorticity it is directly related to 
mean spin vector of spin 1/2 particles with four-momentum $p$, 
produced around point $x$ on freeze-out hypersurface 
   \begin{eqnarray}
\label{xp-pol}
 S^\mu(x,p)
 =\frac{1}{8m}     [1-n_F(x,p)] \: p_\sigma \epsilon^{\mu\nu\rho\sigma} 
  \varpi_{\rho\nu}(x) 
   \end{eqnarray}
where $n_F(x,p)$ is the Fermi-Dirac distribution function and $m$ is mass of the 
considered particle. 
The polarization vector of $S$-spin particle is defined as 
   \begin{eqnarray}
   \label{P_S}
  P^\mu_{S} = S^\mu / S.  
   \end{eqnarray}

The polarization of the $\Lambda$ hyperon is measured in
its rest frame, therefore the $\Lambda$ polarization is 

   \begin{eqnarray}
   \label{P_L-rest}
  P^\mu_{\Lambda} = 2 S^{*\mu}_{\Lambda}  
   \end{eqnarray}
where $S^{*\mu}_{\Lambda}$ is mean spin vector of the $\Lambda$ hyperon in its rest frame. 
The zeroth component  $S^{0}_{\Lambda}$
 identically vanishes in the $\Lambda$ rest frame
and the spatial component becomes \cite{Kolomeitsev:2018svb}
   \begin{eqnarray}
\label{S-rest}
 {\bf S}^*_{\Lambda}(x,p)
 = {\bf S}_{\Lambda} - 
 \frac{{\bf p}_{\Lambda} \cdot {\bf S}_{\Lambda}}{E_{\Lambda}(E_{\Lambda}+m_{\Lambda})}
 {\bf p}_{\Lambda}, 
   \end{eqnarray}
where $E_\Lambda=\sqrt{m_\Lambda^2 + {\bf p}^2}$.
Substitution of the  expression for ${\bf S}$ from Eq. (\ref{xp-pol}) and averaging this 
expression over the ${\bf p}_{\Lambda}$ direction (i.e. over ${\bf n}_p$) 
results in the following 
polarization in the direction orthogonal to the reaction plane ($xz$) \cite{Kolomeitsev:2018svb} 
   \begin{eqnarray}
   \label{P_Lambda}
\langle  P_{\Lambda}\rangle_{{\bf n}_p} =  
 \frac{1}{2m_{\Lambda}}
 \left(E_{\Lambda} - \frac{1}{3} \frac{{\bf p}_{\Lambda}^2}{E_{\Lambda}+m_{\Lambda}} \right)
 \varpi_{zx},  
   \end{eqnarray}
where $m_{\Lambda}$  is the $\Lambda$ mass, 
$E_{\Lambda}$ and ${\bf p}_{\Lambda}$ are the energy and momentum of the emitted $\Lambda$ hyperon, respectively. 
Here we put $(1-n_\Lambda) \simeq 1$ because the $\Lambda$ production takes place only 
in high-temperature regions, where Boltzmann statistics dominates.

Particles are produced across entire
freeze-out hypersurface. Therefore to calculate the global
polarization vector, the above
expression should be averaged over the freeze-out hypersurface $\Sigma$
and particle momenta 
   \begin{eqnarray}
\label{polint}
 P_{\Lambda}^\varpi
 = \frac{\int (d^3 p/p^0) \int_\Sigma d \Sigma_\lambda p^\lambda
n_{\Lambda}  \langle  P_{\Lambda}\rangle_{{\bf n}_p}}
 {\int (d^3 p/p^0) \int_\Sigma d\Sigma_\lambda p^\lambda \, n_{\Lambda}}. 
   \end{eqnarray}
Here $P_{\Lambda}$ is averaged over the whole system and momenta of emitted particles. 
Application of the experimental rapidity acceptance is performed in terms of a so-called 
hydrodynamical rapidity 
   \begin{eqnarray}
   \label{y}
y_h = \frac{1}{2} \ln \frac{u^0+u^3}{u^0-u^3} , 
   \end{eqnarray}
based on hydrodynamical 4-velocity $u^\mu$. The  $d\Sigma_\lambda p^\lambda$ integration runs 
only over those cells, where condition $|y_h|<y_{\rm{acceptance}}$ is met. 
Let us denote this restricted freeze-out hypersurface as $\Sigma_{\Delta y}$.  
Of course, this is only imitation of the actual experimental acceptance. 
Unfortunately, imitation of 
transverse-momentum acceptance in the similar manner is impossible because  
the transverse momentum is mainly determined by thermal motion in the cell.

\begin{figure}[htb]
\includegraphics[width=6.6cm]{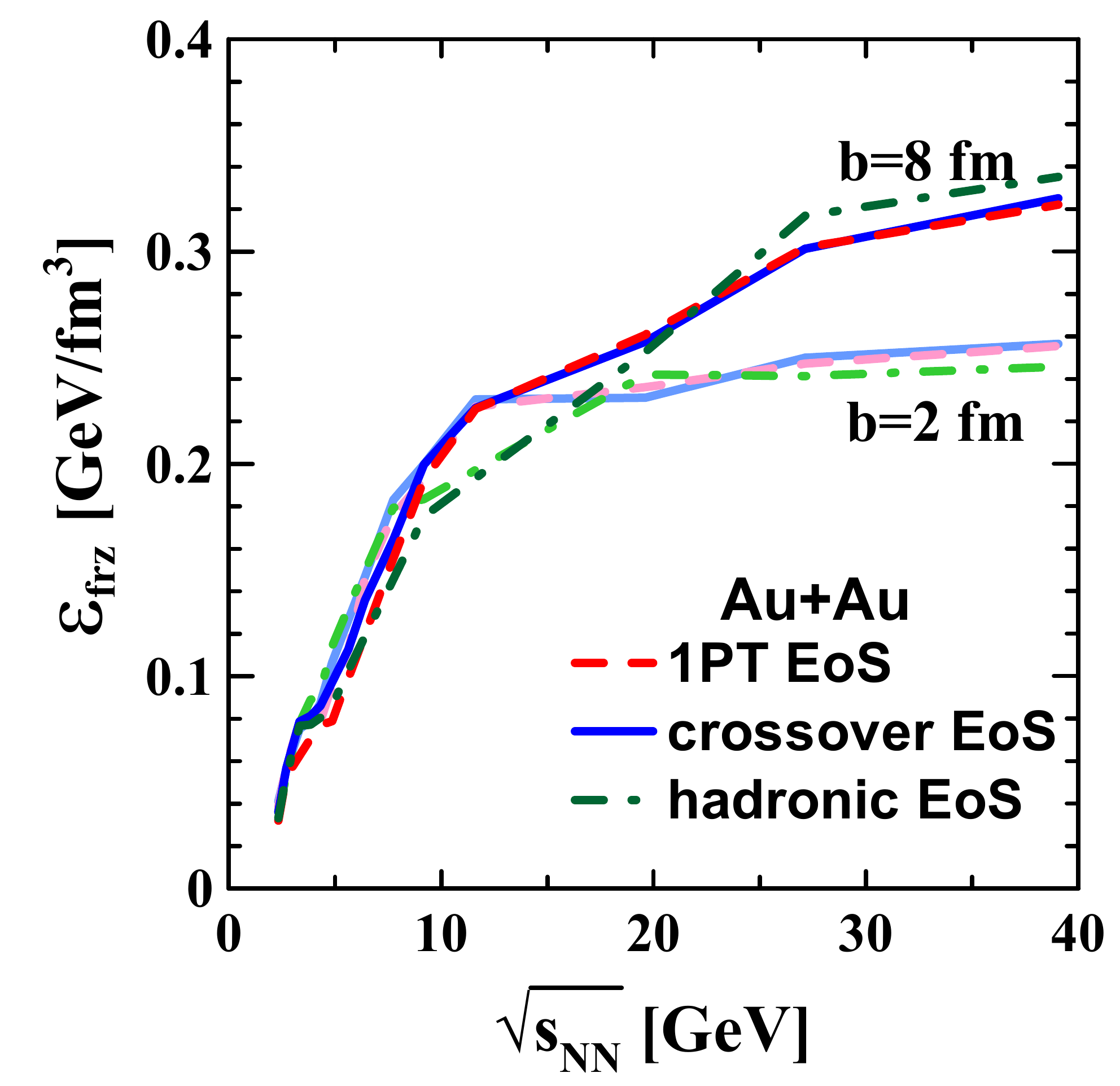}
 \caption{(Color online)
Average actual freeze-out energy density versus collision energy $\sqrt{s_{NN}}$
in Au+Au collisions at impact parameters $b=$ 2 and 8 fm 
calculated with different EoS's. Lower set of (pale) lines corresponds to $b=$ 2 fm.  
}
\label{fig1}
\end{figure}

Similarly to previous 3FD simulations \cite{Ivanov:2020udj,Ivanov:2019ern,Ivanov:2019wzg,Ivanov:2020wak},
a simplified version of the freeze-out is used. 
The freeze-out is isochronous that, in particular, implies 
$(d^3 p/p^0) d\Sigma_\lambda p^\lambda =d^3 p \;d^3 x $. 
The freeze-out instant is associated with time, when the energy density 
$\langle \varepsilon (t)\rangle$ averaged over the central region (i.e. slab $|z|\leq$ 4 fm)
reaches the value of the average freeze-out energy density in the same central region
obtained in conventional 3FD simulations with differential, i.e.
cell-by-cell, freeze-out \cite{Russkikh:2006aa,Ivanov:2008zi}. 
This actual freeze-out energy density, $\varepsilon_{\scr{frz}}$, averaged over frozen out system, 
is illustrated in Fig. \ref{fig1} for two impact parameters and different EoS's. 
It is important to note that values of $\varepsilon_{\scr{frz}}$ in Fig. \ref{fig1} are not paramrters 
of the 3FD model. They are automatically generated in the 3FD simulations as a result of 
the implemented freeze-out dynamics described in Refs. \cite{Russkikh:2006aa,Ivanov:2008zi}.
The only freeze-out parameter is $\epsilon_{\scr{frz}}= 0.4$ GeV/fm$^3$, which 
has a meaning of a ``trigger'' energy density, at which the freeze-out procedure starts. 
This parameter is the same for all EoS's and all collision energies.

We can simplify Eq. (\ref{polint}) by explicitly performing integration over $dp$. 
We reorganize terms in parentheses in Eq. (\ref{P_Lambda}) and use the following relations
   \begin{eqnarray}
   \label{rhoH}
\int d^3 p \; d^3 x \; n_{\Lambda} &=& \int d^3 x \; \rho_{\Lambda} \\
   \label{T00H}
\int  d^3 p \; d^3 x\; E_{\Lambda} \; n_{\Lambda} &=& \int d^3 x \;  T^{00}_{\Lambda} 
   \end{eqnarray}
where $\rho_{\Lambda}$ is the $\Lambda$ density in the frame of calculation and 
$T^{00}_{\Lambda}$ is the $00$ component of the partial energy-momentum tensor related to 
the $\Lambda$ contribution 
   \begin{eqnarray}
   \label{T00Heq}
T^{00}_{\Lambda} = (\varepsilon_{\Lambda} + p_{\Lambda})u^0 u^0 - p_{\Lambda}
   \end{eqnarray}
with $\varepsilon_{\Lambda}$ and $p_{\Lambda}$ being the corresponding partial 
energy density and pressure, respectively. 
$\rho_{\Lambda}$, $\varepsilon_{\Lambda}$ and $p_{\Lambda}$ are determined by ideal-gas 
relations in terms of temperature, baryon and strange chemical potentials. Note that 
the system is described by the ideal-gas EoS at the freeze-out stage.
Thus, inserting expression (\ref{P_Lambda}) for
$\langle  P_{\Lambda}\rangle_{{\bf n}_p}$ into Eq. (\ref{polint}) and performing the 
above described manipulations we arrive at 
   \begin{eqnarray}
\label{polint-fin}
 P_{\Lambda}^\varpi
 = \frac{1}{6} \frac{\displaystyle\int_{\Sigma_{\Delta y}} d^3 x 
 \left(\rho_{\Lambda} +  2 T^{00}_{\Lambda}/m_{\Lambda} \right) 
\varpi_{zx}}%
{\displaystyle\int_{\Sigma_{\Delta y}}  d^3 x \, \rho_{\Lambda}}. 
   \end{eqnarray}
This is the final expression with which we perform our simulations.

In previous calculations \cite{Ivanov:2020udj,Ivanov:2019ern,Ivanov:2019wzg,Ivanov:2020wak}, 
the $n_{\Lambda}$ weight in Eq. (\ref{polint}) was replaced by the energy-density weight. 
Moreover, averaging of $\varpi_{zx}$ and the term in parentheses in Eq. (\ref{P_Lambda}) was decoupled. 
In the present approach we avoid these approximations.

\subsection{Polarization transfer in two-body decays}
\label{Polarization transfer}

Only a fraction of all
detected $\Lambda$'s are produced directly at the 
freeze-out stage. These are primary $\Lambda$'s. A 
fraction of $\Lambda$'s originates from decays of heavier hyperons. 
The most important feed-down channels are strong decays of 
$\Sigma^* \rightarrow \Lambda + \pi$ and electromagnetic decays 
$\Sigma^0 \rightarrow \Lambda + \gamma$. 
When polarized particles decay, their daughters are
themselves polarized because of angular momentum conservation. 
The amount of polarization that is transferred to
the daughter particle depends on the momentum
of the daughter in the rest frame of the parent. 
For the mean, momentum-integrated,
spin vector in the rest frame, a simple linear rule applies
   \begin{eqnarray}
\label{transfer}
{\bf S}_D^* = C {\bf S}_P^* ,  
\end{eqnarray}
where $P$ is the parent particle, $D$ is the daughter and $C$ is 
a coefficient, values of which are presented in Table I of Ref. \cite{Becattini:2016gvu}. 
Making use of these $C$ coefficients, we arrive at the following expression for 
the observable $\Lambda$ polarization
   \begin{eqnarray}
\label{polint-fin-decays}
 (P_{\Lambda}^\varpi)_{\rm{obs.}}
 = \frac{N_{\Lambda} P_{\Lambda}^\varpi + (5/3) N_{\Sigma^*} P_{\Sigma^*}^\varpi-
(1/3) N_{\Sigma^0} P_{\Sigma^0}^\varpi}%
{N_{\Lambda} +  N_{\Sigma^*} + N_{\Sigma^0}},   
   \end{eqnarray}
where $P_{Y}^\varpi$ is the global polarization $Y$ hyperon ($Y=\Lambda,\Sigma^*,\Sigma^0$) 
calculated similarly to Eq. (\ref{polint-fin}) and 
   \begin{eqnarray}
\label{N_Y}
N_{Y} = \int_{\Sigma_{\Delta y}}  d^3 x \, \rho_{Y}
   \end{eqnarray}
is the total number of $Y$ hyperons ($Y=\Lambda,\Sigma^*,\Sigma^0$) 
on the freeze-out hypersurface $\Sigma_{\Delta y}$.
In Eq. (\ref{N_Y}) we neglected contribution of the decay channel 
$\Sigma^* \rightarrow \Sigma + \pi$ with small branching ratio (0.117), 
and hence put the branching ratio of the $\Sigma^* \rightarrow \Lambda + \pi$
channel equal to unit. 

%
\begin{figure}[bht]
\includegraphics[width=8.5cm]{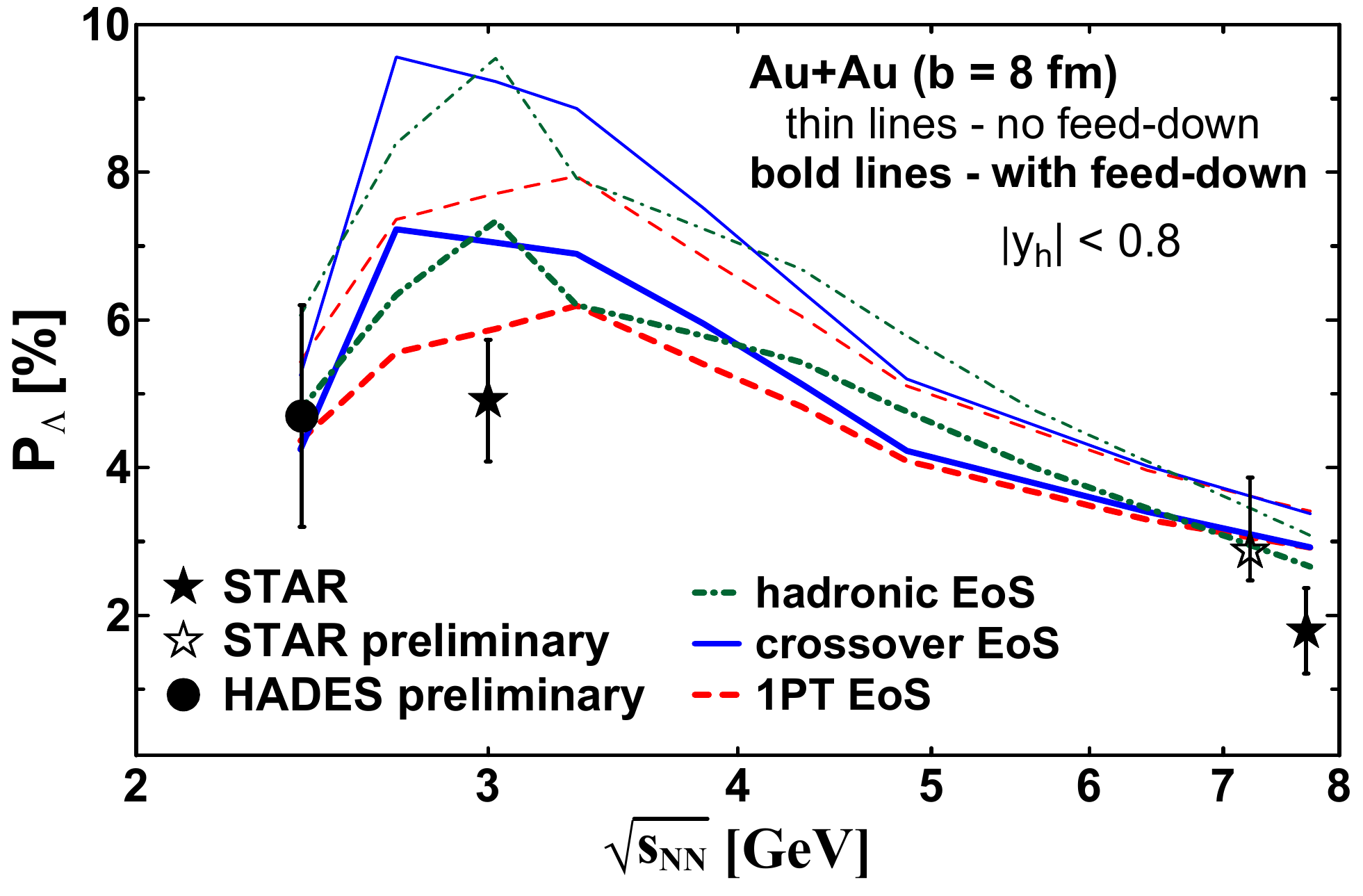}
 \caption{(Color online)
 Global $\Lambda$ polarization in midrapidity region ($|y_h|<0.8$), 
originated from the thermal vorticity, in Au+Au collisions at $b=$ 8 fm as 
function of collision energy $\sqrt{s_{NN}}$ calculated with different EoS's.
Result with contribution of the feed-down from higher-lying resonances 
(bold lines) and without them (thin lines).    
 Data are from Refs. \cite{STAR:2017ckg,STAR:2021beb,Okubo:2021dbt} (STAR) 
and \cite{HADES:SQM2021} (HADES). 
}
\label{fig2}
\end{figure}

In Fig. \ref{fig2} we demonstrate the effect of the feed-down from higher-lying resonances on 
the global $\Lambda$ polarization in midrapidity region ($|y_h|<0.8$). 
The impact parameter $b=$ 8 fm roughly 
roughly comply with the STAR centrality selection of
20--50\% \cite{STAR:2017ckg}. 
To associate these impact parameters with collision centrality, one should 
keep in mind  that in the 3FD model the colliding nuclei have a shape of sharp spheres without 
the Woods-Saxon diffuse edge. 
This fact, implemented in the Glauber simulations 
based on the nuclear overlap calculator \cite{web-docs.gsi.de},   
results in this mean impact parameter, which is shifted  
by $\approx$1.5 fm down,  as compared the result of Ref. \cite{Abelev:2008ab}. 
The width of this 
midrapidity region is chosen on the condition of the best reproduction of the STAR acceptance  
$|\eta|<1$ \cite{STAR:2017ckg}, where $\eta$ is pseudorapidity. 
This window is not that good for the low-energy data \cite{STAR:2021beb,Okubo:2021dbt,HADES:SQM2021}, 
where rapidity acceptance is asymmetric with respect to the midrapidity. However, it is still good 
in view of flat rapidity dependence of the observed $P_\Lambda$. Results are presented for three EoS's. 
As seen, the feed-down reduces $P_\Lambda$ by $\approx$25\% at low energies 
and by $\approx$15\% at 7.7 GeV.   
All results presented below are calculated taking into account feed-down from higher-lying resonances.

\section{Meson-field induced polarization}
\label{Meson-field}

In this section, meson-field induced contribution to the global polarization is discussed. 
It was proposed in Ref. \cite{Csernai:2018yok} primarily to explain the observed 
$\Lambda$-$\bar{\Lambda}$ splitting in the global polarization. We do not discuss this splitting 
in the present paper, because it deserves special separate discussion, but rather study the effect 
of the meson-field induced contribution on the $\Lambda$ polarization. 
Below, we briefly repeat derivation of Refs. \cite{Csernai:2018yok,Xie:2019wxz} with the same result 
but somewhat different reasoning. 

Let strong interaction among baryons be mediated by a scalar field $\sigma$ and a vector field $V^{\mu}$,  
as it assumed in the Walecka model \cite{Walecka74,Serot-Walecka}. 
The effective Lagrangian of this model is
\begin{eqnarray}
\lefteqn{{\cal L}_{\rm eff} = \sum_j \bar{\psi}_j ( i \! \not\!\partial - m_{j} 
+ g_{\sigma j} \sigma - g_{V j} \not\!V ) \psi_j} \nonumber 
\\
&& \hspace*{-9mm}
+ \thalf \left( \partial_{\mu} \sigma \partial^{\mu} \sigma
- m_{\sigma}^2 \sigma^2 \right)
 - \oneqt V^{\mu\nu}V_{\mu\nu}
+ \thalf m_{V}^2 V_{\mu}V^{\mu} \,.
\label{Leff}
\end{eqnarray}
Here $j$ labels the spin-1/2 baryons, and the field strength tensor for the vector field is  
\begin{eqnarray}
V_{\mu\nu} = \partial_\mu V_\nu - \partial_\nu  V_\mu \,.
\label{e1}
\end{eqnarray}
In general, the Lagrangian may include a potential $U(\sigma)$ of $\sigma$-field self-interaction, 
but its exact form is irrelevant here. Therefore, we put $U(\sigma)=0$ for definiteness. 
The $V$ field is usually associated with the vector $\omega$ meson and $\sigma$ -- with the $\sigma$ meson. 
The $\sigma$ field results in an attractive interaction and $\omega$, a repulsive interaction.  
$g_{\omega j}$ and $g_{\sigma j}$ are the coupling constants, possible values of which can be found, e.g., 
in Ref.  \cite{Kapusta-Gale}.

The $\sigma$  and $\omega$ fields are treated in the mean field approximation 
\cite{Walecka74,Serot-Walecka,Kapusta-Gale}: 
\begin{eqnarray}
\partial^2 V^{\nu} + m^2_{V} V^\nu = \sum_j g_{V j} J^\nu_j \,.
\label{field_omega}
\end{eqnarray}
where $J^\mu_j=\langle \bar{\psi} \gamma^\mu \psi\rangle$ is the baryon current of $j$ baryons, 
in which baryons and antibaryons contribute with opposite signs, and   
\begin{eqnarray}
\partial^2 \sigma + m_{\sigma}^2 \sigma = \sum_j g_{\sigma j} n_{sj}
\label{field_sigma}
\end{eqnarray}
where $n_{sj}=\langle \bar{\psi} \psi\rangle$ is the scalar density, in which 
baryons and antibaryons contribute with the same signs.
It is expected that these interactions in terms of hadrons 
are relevant at the freeze-out stage even if the preceding evolution was dominated by the quark-gluon phase. 
At this stage 
the corresponding energy scale is much less than $m_\omega = 783$ MeV and $m_\sigma \approx 600$ MeV. 
Therefore, the derivatives in Eqs. (\ref{field_omega}) and (\ref{field_sigma}) can be neglected, and thus 
we arrive at the following solution for the fields
\begin{eqnarray}
\sigma &=& \frac{1}{m_{\sigma}^2}\sum_j g_{\sigma j} n_{sj}, 
\\
 V^\nu &=& \frac{1}{m^2_{V}}\sum_j g_{V j} J^\nu_j \simeq \frac{\bar{g}_{V}}{m^2_{V}} J^\nu_B. 
\label{sigma-omega-sol}
\end{eqnarray}
The $V^\nu$ field can be approximately expressed through the baryon current 
$J^\nu_B=n_B u^\nu$, where $n_B$ is the baryon density and 
$\bar{g}_{V}$ is the mean coupling constant of the vector meson.

Nonrelativistic reduction of the interaction between the fields and the spin operator 
$\widehat{\bf S}$ of the $\Lambda$ and $\bar{\Lambda}$ hyperons is performed by means of  
the Foldy-Wouthuysen transformation 
\cite{FW,BD,Greiner}, i.e. an expansion in powers of the inverse of baryon masses,  
which complies with neglecting derivatives in Eqs. (\ref{field_omega}) and (\ref{field_sigma}). 
The nonrelativistic interaction of the spin with the meson fields reads 
\begin{eqnarray}
\widehat{H}_{\rm spin} &=&
 \frac{g_{\sigma\Lambda}}{2 m^2_\Lambda} \widehat{\bf S} \cdot {\bf \nabla} \sigma \times \widehat{\bf p}
-\frac{g_{V\Lambda}}{m_\Lambda} \beta 
\widehat{\bf S} \cdot {\bf B}_V 
\cr  
&-& i \frac{g_{V\Lambda}}{4 m^2_\Lambda}
\widehat{\bf S} \cdot {\bf \nabla}\times  {\bf E}_V 
- \frac{g_{V\Lambda}}{2 m^2_\Lambda}
\widehat{\bf S} \cdot {\bf E}_V \times \widehat{\bf p}. 
\label{e2}
\end{eqnarray}
Here 
${\bf E}_V$ and  ${\bf B}_V$ are the vector-meson electric and magnetic fields 
\begin{eqnarray}
E_i &= & V_{i0}, 
\label{BI1} 
\\
B_i &= & -\frac12 \varepsilon_{ijk} V^{jk},
\label{BI2}
\end{eqnarray}
where $i, j, k =1, 2, 3$, 
$\widehat{\bf p}$ is the momentum operator of the $\Lambda$ or $\bar{\Lambda}$, and 
\begin{eqnarray}
\beta = 
\left( \begin{array}{cc}
		1 & 0 \\ 0 & -1
\end{array} \right)
\label{e3}
\end{eqnarray}
is the usual Dirac $4 \times 4$ $\beta$ matrix, resulting in opposite signs 
when acting on the $\Lambda$ and $\bar{\Lambda}$ spinors.

Let us turn to the density operator
\begin{eqnarray}
\widehat{\rho}
  = \frac{1}{Z} \exp[-\widehat{H}/T  
	+ \nu \widehat{Q}/T +\boldsymbol{\omega} \cdot (\widehat{\bf L} + 
  \widehat{\bf S})/T]  
	\nonumber \\
\label{rho_tot}
\end{eqnarray}
where $\widehat{H}$ is the Hamiltonian, $T$ is the temperature, 
$\widehat{Q}$ stands for conserved charges (baryon, electric, strangeness) with
$\nu$ being the corresponding chemical potentials. 
The angular
velocity $\boldsymbol{\omega}$ plays the role of a chemical potential for the
angular momentum, consisting of the orbital ($\widehat{\bf L}$) and spin ($\widetilde{\bf S}$) parts.

Inspecting the spin-dependent part of the Hamiltonian, Eq. (\ref{e2}), 
we see that only the second and third terms on the r.h.s. can be associated with 
additional corrections to the spin chemical potential, provided the equilibrium is 
local.  
The first and forth terms also 
produce the polarization, but a chaotic one, because its direction depends on 
the momentum direction. However, they may induce a collective polarization, 
if there is a strong collective flow, i.e. if particle momenta are dominantly aligned 
along certain direction. This polarization would be similar to that discussed in Refs. 
\cite{Ivanov:2018eej,Xia:2018tes,Lisa:2021zkj}. 
The third term contains the extra derivative 
in the nominator and the extra $\Lambda$ mass ($m_\Lambda$) in the denominator, as  
compared to the second term. This combination amounts to a smallness parameter, 
which has been already used when neglecting derivatives in the mean field equations  
(\ref{field_omega}) and (\ref{field_sigma}). Besides, only the sum 
of the third and fourth terms in Eq. (\ref{e2}) is Hermitian, not the individual terms. 
Therefore, it is reasonable to disregard them together. 
Thus, we can represent the density operator relevant to the global polarization in the following form 
\begin{eqnarray}
\widehat{\rho}
  = \frac{1}{Z} \exp\left[-\frac{\widehat{\widetilde{H}}}{T} + \frac{\nu}{T} \widehat{Q} 
	+\frac{g_{V\Lambda}}{m_\Lambda T} \beta \widehat{\bf S} \cdot \boldsymbol{B}_V  
	+\frac{\boldsymbol{\omega}}{T} \cdot (\widehat{\bf L} +   \widehat{\bf S})\right]  
	\nonumber \\
\label{rho_s}
\end{eqnarray}
where the term with the extra spin chemical potential from $\widehat{H}_{\rm spin}$ is explicitly displayed, 
while $\widehat{\widetilde{H}}$ denotes the rest part of the Hamiltonian.

Derivation along the lines of Ref. \cite{Becattini:2007nd} results in 
the mean spin vector of the hyperons 
($Y = \Lambda$ or $\bar{\Lambda}$) 
with four-momentum $p$, produced around point $x$ 
\begin{eqnarray}
\label{pixpgen3}
   S_Y^\mu(x,p) = \frac{1}{4} \left( \varpi_{\rm c}^\mu + 
	\beta_Y \frac{g_{V\Lambda}}{m_\Lambda T} B_{V}^\mu \right)
\end{eqnarray}
where $\beta_\Lambda =1$   and $\beta_{\bar{\Lambda}}=-1$, 
\begin{eqnarray}
\label{varpi_c}
\varpi_{\rm c}^\mu = -\frac{1}{2} \epsilon^{\mu\rho\sigma\tau} \varpi_{\rho\sigma} p_\tau/m_\Lambda	
\end{eqnarray}
is the comoving axial thermal vorticity defined in terms of the thermal vorticity (\ref{therm.vort.}).
Here we returned to the relativistic treatment of the rotation, therefore the 
angular velocity $\boldsymbol{\omega}$ was replaced by the relativistic thermal vorticity. 
We can make expression (\ref{pixpgen3}) completely covariant by identifying the 
magnetic field $B_{V}^\mu$ with the comoving magnetic field
\begin{equation}
\label{B_c}
B_{V}^\mu = B_{c}^\mu = -\frac{1}{2}
\epsilon^{\mu\rho\sigma\tau} V_{\rho\sigma} p_\tau/m_\Lambda.  
\end{equation}
This is a certain ansatz because the corresponding interaction (\ref{e2}) was originally derived 
in the nonrelativistic approximation.  
In explicitly covariant form Eq. (\ref{pixpgen3}) reads
\begin{eqnarray}
\label{pixpgen2}
   S_Y^\mu(x,p) = - \frac{1}{8m_\Lambda} \epsilon^{\mu\rho\sigma\tau} p_\tau 
  \left( \varpi_{\rho\sigma} + \beta_Y \frac{g_{V\Lambda}}{m_\Lambda T} V_{\rho\sigma} \right).
\end{eqnarray}
The vector meson field enters this expression similarly to the electromagnetic field interacting 
with magnetic moment of the $Y$-hyperon \cite{Becattini:2016gvu}. 
This expression is valid \cite{Becattini:2007nd} (see also Ref. \cite{Becattini:2016gvu})  
in the leading order in the thermal vorticity and field strength tensor. The Fermi factor 
$[1-n_Y(x,p)]$ was again omitted because of its negligible effect at high temperatures achieved in nuclear collisions.

The further derivation is identical to that performed in the previous section with the substitution 
$\left( \varpi_{\rho\sigma} + \beta_Y \frac{g_{V\Lambda}}{m_\Lambda T} V_{\rho\sigma} \right)$ 
instead of $\varpi_{\rho\sigma}$. Finally we arrive to the following expression for the 
meson-field contribution to the global polarization of the $Y$ hyperon ($Y=\Lambda$ or $\bar{\Lambda}$): 
   \begin{eqnarray}
\label{polint-V-fin}
 P_{Y}^V
 = 
\frac{\beta_Y g_{V\Lambda}}{6m_\Lambda T}
\frac{\displaystyle\int_{\Sigma_{\Delta y}} d^3 x 
 \left(\rho_{Y} +  2 T^{00}_{Y}/m_{\Lambda} \right) V_{zx}}%
{\displaystyle\int_{\Sigma_{\Delta y}}  d^3 x \, \rho_{Y}}. 
   \end{eqnarray}
which should be added to the thermal-vorticity term (\ref{polint-fin}). 
Here $\beta_\Lambda =1$   and $\beta_{\bar{\Lambda}}=-1$, and $V_{zx}$
is defined in terms of the baryon current, $J^\nu_B$, by Eq. (\ref{sigma-omega-sol}). 
The feed-down correction (\ref{polint-fin-decays}) should be applied to the sum of 
thermal-vorticity and meson-field terms.

%
\begin{figure}[bht]
\includegraphics[width=8.5cm]{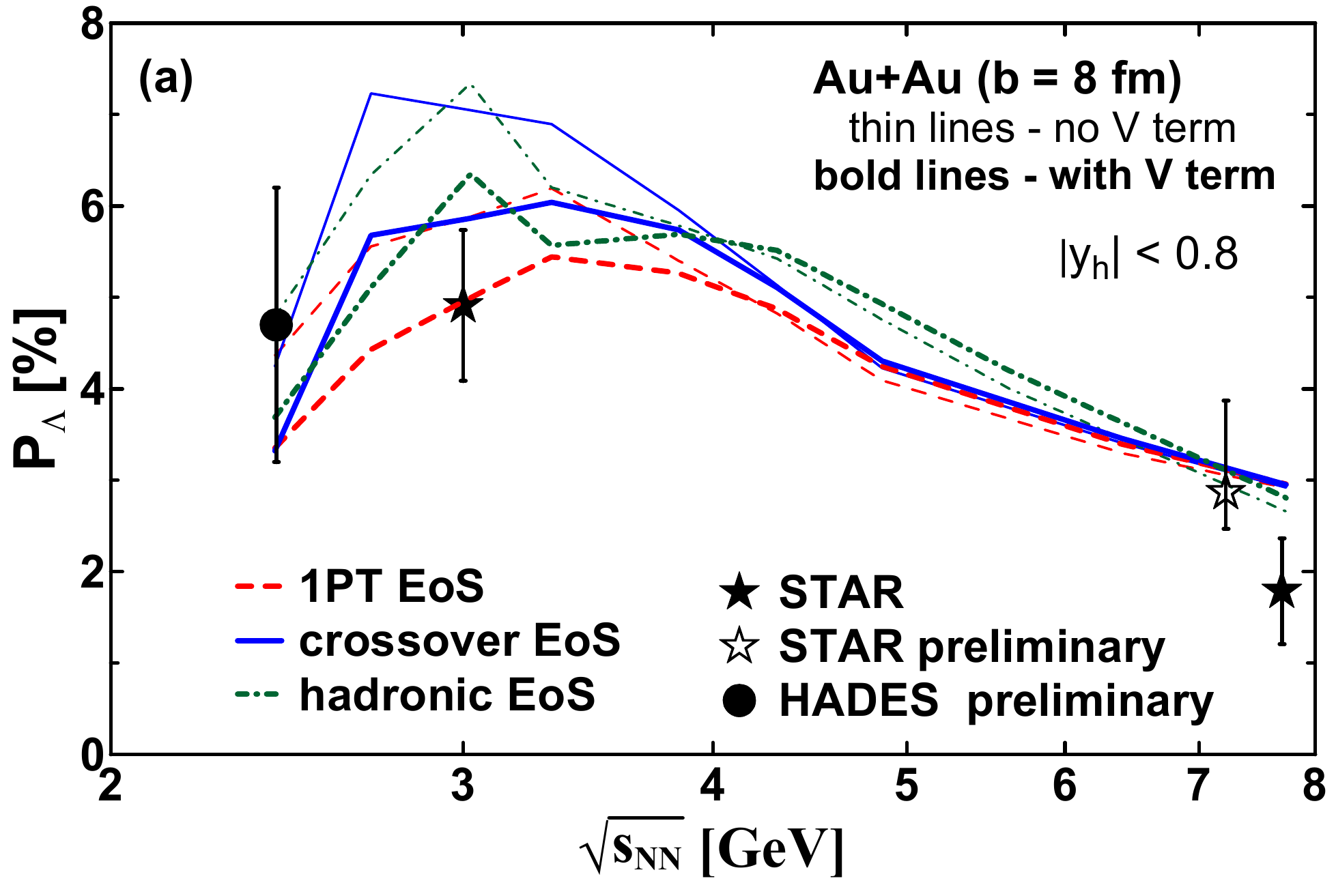}
\includegraphics[width=8.5cm]{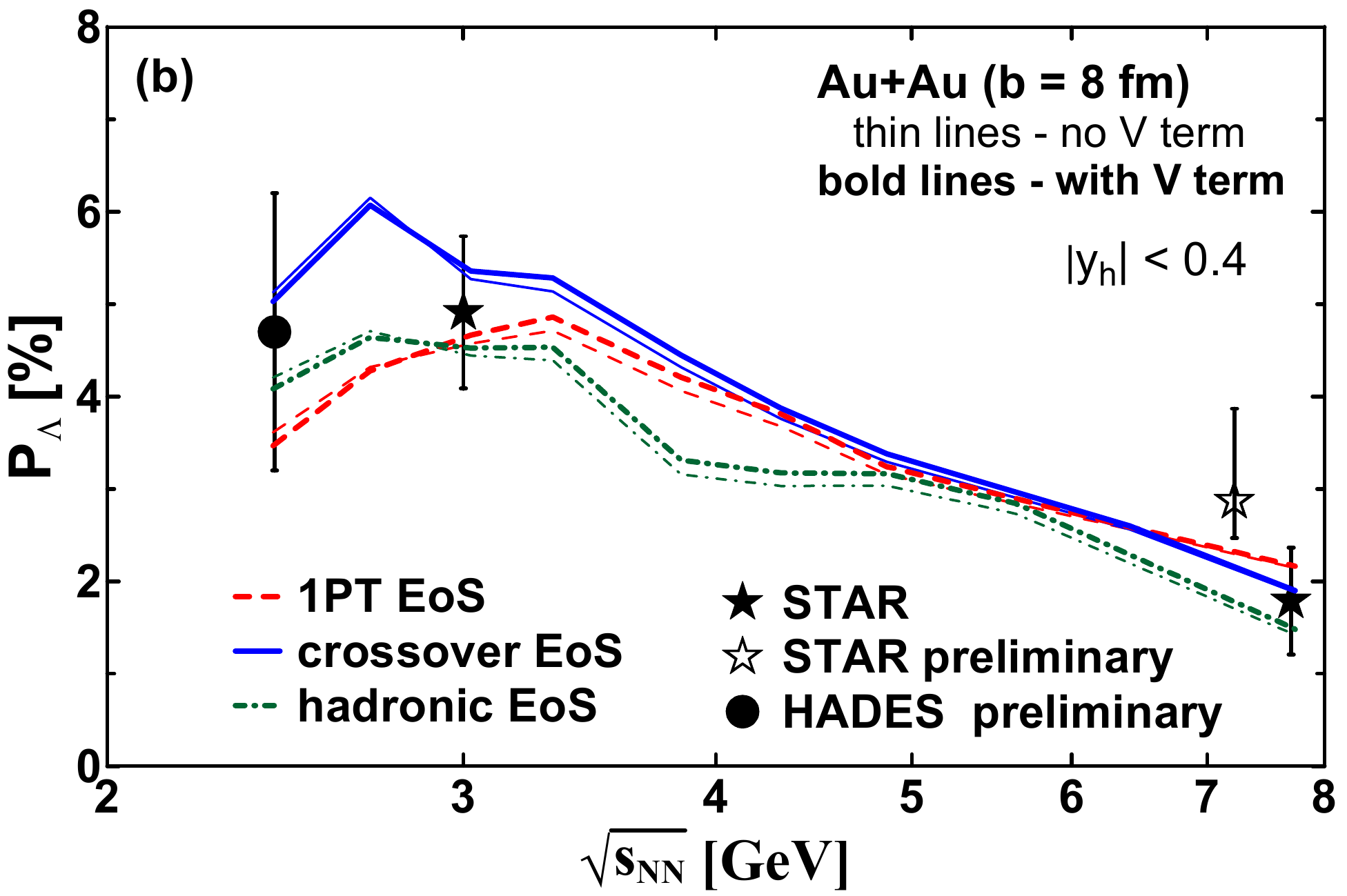}
 \caption{(Color online)
 Global $\Lambda$ polarization in midrapidity regions $|y_h|<0.8$ (a) 
and $|y_h|<0.4$ (b), 
originated from the thermal vorticity with (bold lines) and without (thin lines) the meson-field contribution, 
in Au+Au collisions at $b=$ 8 fm as 
function of collision energy $\sqrt{s_{NN}}$.
Results for different EoS's are presented.
 Data are from Refs. \cite{STAR:2017ckg,STAR:2021beb,Okubo:2021dbt} (STAR) 
and \cite{HADES:SQM2021} (HADES). 
}
\label{fig3}
\end{figure}
%
%
\begin{figure*}[htb]
\includegraphics[width=17.7cm]{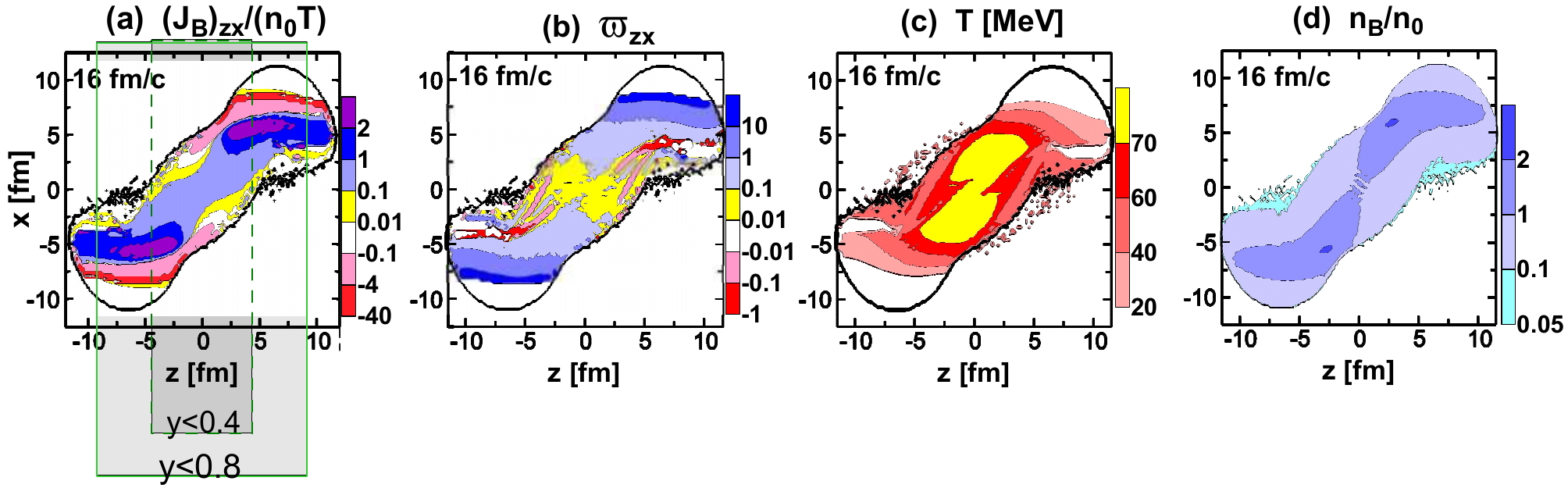}
 \caption{(Color online)
 Panels from left to right: 
(a) $J^B_{zx}/(n_0 T)$, the proper-energy-density weighted relativistic baryon-current $zx$ vorticity, 
Eq. (\ref{JB_munu}), divided by temperature ($T$) and normal nuclear density ($n_0$), 
(b) $\varpi_{zx}$, the similarly weighted thermal $zx$ vorticity, 
(c) $T$, the temperature, and 
(d) $n_B/n_0$, the proper baryon density in units of the normal nuclear density ($n_0$) 
in the reaction plane ($xz$) at time instant $t=$ 16 fm/c
in the semi-central ($b =$ 8 fm) Au+Au collision at $\sqrt{s_{NN}}=$ 2.7 GeV. 
Calculations are done with the crossover EoS. $z$ axis is the beam direction. 
Grey-shaded boxes in the $J^B_{zx}/(n_0 T)$ panel indicate approximate borders of the 
midrapidity regions $|y_h|<0.8$ (light-gray outer box)  
and $|y_h|<0.4$ (dark-gray inner box), where $y_h$ is the hydrodynamical rapidity, see Eq. (\ref{y}). 
}
\label{fig3v}
\end{figure*}

For practical calculations the coupling constant $g_{V\Lambda}$ and 
the mean coupling constant $\bar{g}_{V}$ of the vector meson are needed.
A brief survey of various parametrizations of the relativistic mean-field (RMF) model
is presented in Ref. \cite{Csernai:2018yok}, see also \cite{Weissenborn:2011ut,Maslov:2015wba}.
We use just one of the possible parametrizations: 
$\bar{g}_{V}=g_{VN}=$ 8.646, $g_{V\Lambda}=0.67 g_{VN}$ and $m_V=m_\omega=$ 783 MeV \cite{Cohen:1991qi}. 
The mean coupling constant is associated with the nucleon one because nucleons 
dominate in the baryonic content of system at low energies considered here. 
The uncertainty in the RMF-model parametrization results in the corresponding 
uncertainty in the $P_{Y}^V$ calculation. 

To estimate the scale of the additional $V$-term, we present the terms in parentheses in Eq. (\ref{pixpgen2})
as follows
\begin{eqnarray}
\label{scale}    
  \varpi_{zx} + \frac{g_{\omega\Lambda}}{m_\Lambda T} V_{zx} =
   \varpi_{zx} + \left(\frac{g_{\omega\Lambda}\bar{g}_{\omega} n_0}{m_\Lambda m^2_{\omega}}\right)
	\frac{J^B_{zx}}{T n_0},   
\end{eqnarray} 
where $n_0=$ 0.15 fm$-3$ is the normal nuclear density and 
\begin{eqnarray}
J^B_{\mu\nu} = \partial_\mu J^B_\nu - \partial_\nu  J^B_\mu \,
\label{JB_munu}
\end{eqnarray}
is vorticity of the baryon current, see Eq. (\ref{sigma-omega-sol}). 
The factor 
\begin{eqnarray}
\label{V-factor}
\frac{g_{\omega\Lambda}\bar{g}_{\omega} n_0}{m_\Lambda m^2_{\omega}}\approx 0.1
\end{eqnarray}
is a natural scale of the additional $V$-term. In practice, the contribution of the $V$ term can be 
greater (up to several tens of percent) or less (down to several percent) 
or even have the opposite sign, depending on spatial distributions of thermal and 
baryon-current vorticities and values of the baryon density at the freeze-out.

Figure \ref{fig3} demonstrates the effect of the meson-field contribution to the 
global $\Lambda$ polarization. As seen, the additional meson-field term considerably 
reduces the $\Lambda$ polarization in rapidity window $|y_h|<0.8$
at low collision energies and makes it closer to the STAR data at 3 GeV \cite{STAR:2021beb}. 
At the same time this effect is small in narrower window $|y_h|<0.4$. 
This is a result of the aforementioned spatial distributions of thermal and 
baryon-current vorticities. In Fig. \ref{fig3v}, the spatial distributions of 
the proper-energy-density weighted relativistic baryon-current $zx$ vorticity, 
the similarly weighted thermal $zx$ vorticity, the temperature, and 
the proper baryon density 
in the reaction plane ($xz$) at time instant $t=$ 16 fm/c
in the semi-central ($b =$ 8 fm) Au+Au collision at $\sqrt{s_{NN}}=$ 2.7 GeV are presented. 
Calculations are performed with the crossover EoS.
This time instant of $t=$ 16 fm/c is close to the freeze-out time (16.8 fm/c) determined 
by means of the average freeze-out energy density in the central region
obtained in conventional 3FD simulations, see Fig. \ref{fig1}.

Approximate borders of the regions corresponding to restrictions on 
the hydrodynamical rapidity $y_h$, see Eq. (\ref{y}), are displayed by gray boxes in
the $J^B_{zx}/(n_0 T)$ panel of Fig. \ref{fig3v}: $|y_h|<0.8$ by the light-colored box   
and $|y_h|<0.4$ by the dark-colored box). 
As seen, the baryon-current $zx$ vorticity and the  thermal $zx$ one 
achieve highest absolute values at the participant-spectator border.
Moreover, these values are of the opposite sign. 
The near-border absolute value of $J^B_{zx}/(n_0 T)$
exceeds that of $\varpi_{zx}$. 
Panels ($T$) and ($n_B/n_0$) of Fig. \ref{fig3v} demonstrate that gradients of $1/T$ and $n_B/n_0$
also essentially contribute to $\varpi_{zx}$ and $J^B_{zx}/(n_0 T)$, respectively, 
rather than only vortical motion of the matter.

The $|y_h|<0.8$ region almost completely includes the participant-spectator border. 
Therefore, the $(|y_h|<0.8)$-region integrated baryon-current vorticity
(multiplied by 0.1, see (\ref{V-factor})) considerably reduces the $\varpi_{zx}$-polarization. 
The $|y_h|<0.4$ region only slightly overlaps with the the participant-spectator border. 
Hence, the main contribution to the 
meson-field contribution to the global polarization comes from the bulk, 
where  the baryon-current vorticity is quite moderate. 
Therefore, the $V$-correction to the global polarization in the $|y_h|<0.4$ region
is small.

%
\begin{figure}[bht]
\includegraphics[width=8.5cm]{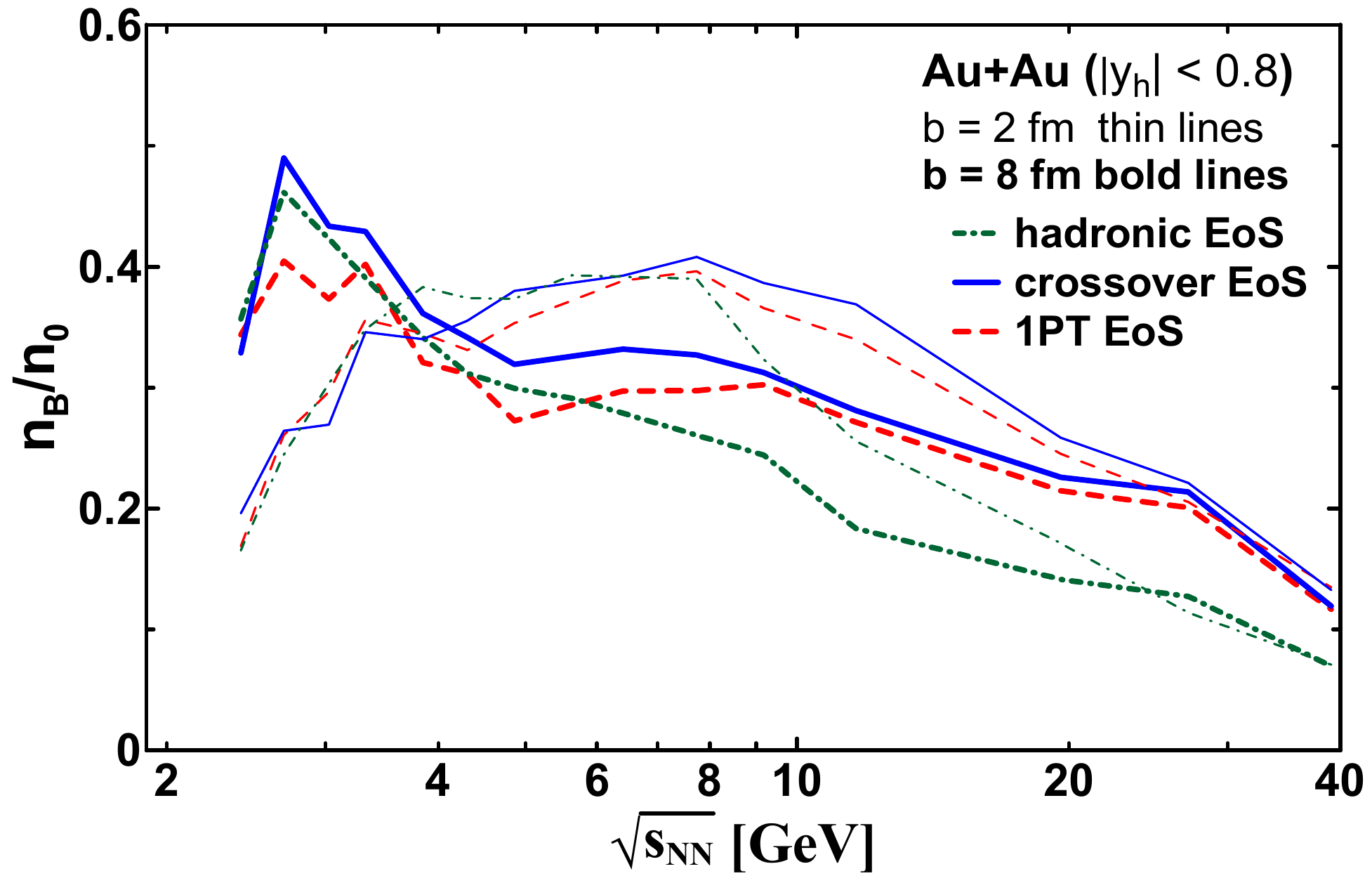}
 \caption{(Color online)
 Mean freeze-out baryon density in units of the normal nuclear density, $n_0=$ 0.15 fm$^{-3}$, 
in midrapidity region ($|y_h|<0.8$) in Au+Au collisions 
at impact parameters $b=$ 8 fm (bold lines) and $b=$ 2 fm  (thin lines) 
as function of collision energy $\sqrt{s_{NN}}$. 
Results for different EoS's are presented.
}
\label{fig3n}
\end{figure}

The effect of the meson-field contribution is negligible at higher energies even in the $(|y_h|<0.8)$ window, 
see Fig. \ref{fig3}. 
The $\Lambda$ polarization  sometimes even increases, 
though only slightly, because of the meson-field contribution. In particular, it 
means that the meson-field induced polarization does not explain the $\Lambda$-$\bar{\Lambda}$
splitting at the energy of 7.7 GeV. This is in contrast to results of Ref. \cite{Xie:2019wxz}, 
where it did explain. The reason is twofold. First, the $(|y_h|<0.8)$ window becomes too narrow 
to cover the participant-spectator border, where 
the baryon-current  vorticity and the  thermal  one 
achieve highest absolute values. Second, the baryon density at the freeze-out decreases  
with the collision energy rise. Indeed, 
the peak value of the baryon density in the $(|y_h|<0.8)$ window at the freeze-out 
occurs precisely at these low collision energies at $b=$ 8 fm, as seen from Fig. \ref{fig3n}. 
This peak is achieved, in particular, because the spectator regions are partially included in this
$(|y_h|<0.8)$ region. 
It seemingly contradicts the results by Cleymans and Randrup \cite{Randrup:2006nr}, 
obtained in the statistical model. They obtained the maximum baryon density at approximately 8 
GeV, when analyzing central collisions. The 3FD model predicts a similar result for the central collisions: 
the maximum $n_B$ is achieved at $\approx$8 GeV at $b=$ 2 fm, see Fig. \ref{fig3n}.

\section{Rapidity dependence}
\label{Rapidity}

The STAR data \cite{STAR:2021beb} on rapidity dependence  of the global $\Lambda$ polarization
at $\sqrt{s_{NN}}=$ 3 GeV are presented for a wide range of centrality selection 0-50\%. 
The nuclear overlap calculator \cite{web-docs.gsi.de}, 
based on the Glauber simulations, predicts the range of impact parameters $b=$ 0--8.8 fm 
for this centrality range. This estimate takes into account  
that the colliding nuclei are sharp spheres without 
the Woods-Saxon diffuse edge in the 3FD model.
Such a wide range cannot be represented by a single impact parameter. 
Therefore, we need to perform averaging over $b$: 
\begin{eqnarray}
\label{PL-mean-def}
 \langle P_\Lambda \rangle
=  \int_0^{b_{max}} b d b \;P_\Lambda (b)\; / \int_0^{b_{max}} b d b
\end{eqnarray}
where $b_{max}=$ 8.8 fm.
Actual 3FD simulations of Au+Au collisions were performed at discrete 
impact parameters $b=$ 2, 4, 6, 8 and 11 fm. Therefore, we replace the integral in Eq. (\ref{PL-mean-def}) 
by a sum over impact parameters
\begin{eqnarray}
\label{PL-mean}
\!\!\!\!\! \langle P_\Lambda \rangle \approx
\sum_{b_i=\rm{2,4,6,8 fm}} \!\!\!\! b_i  \; P_\Lambda (b_i)\; / 
\sum_{b_i=\rm{2,4,6,8 fm}} \!\!\!\! b_i, 
\end{eqnarray}
where $\Delta b$ is canceled because $b_i$ points are equidistant.

%
\begin{figure}[bht]
\includegraphics[width=8.5cm]{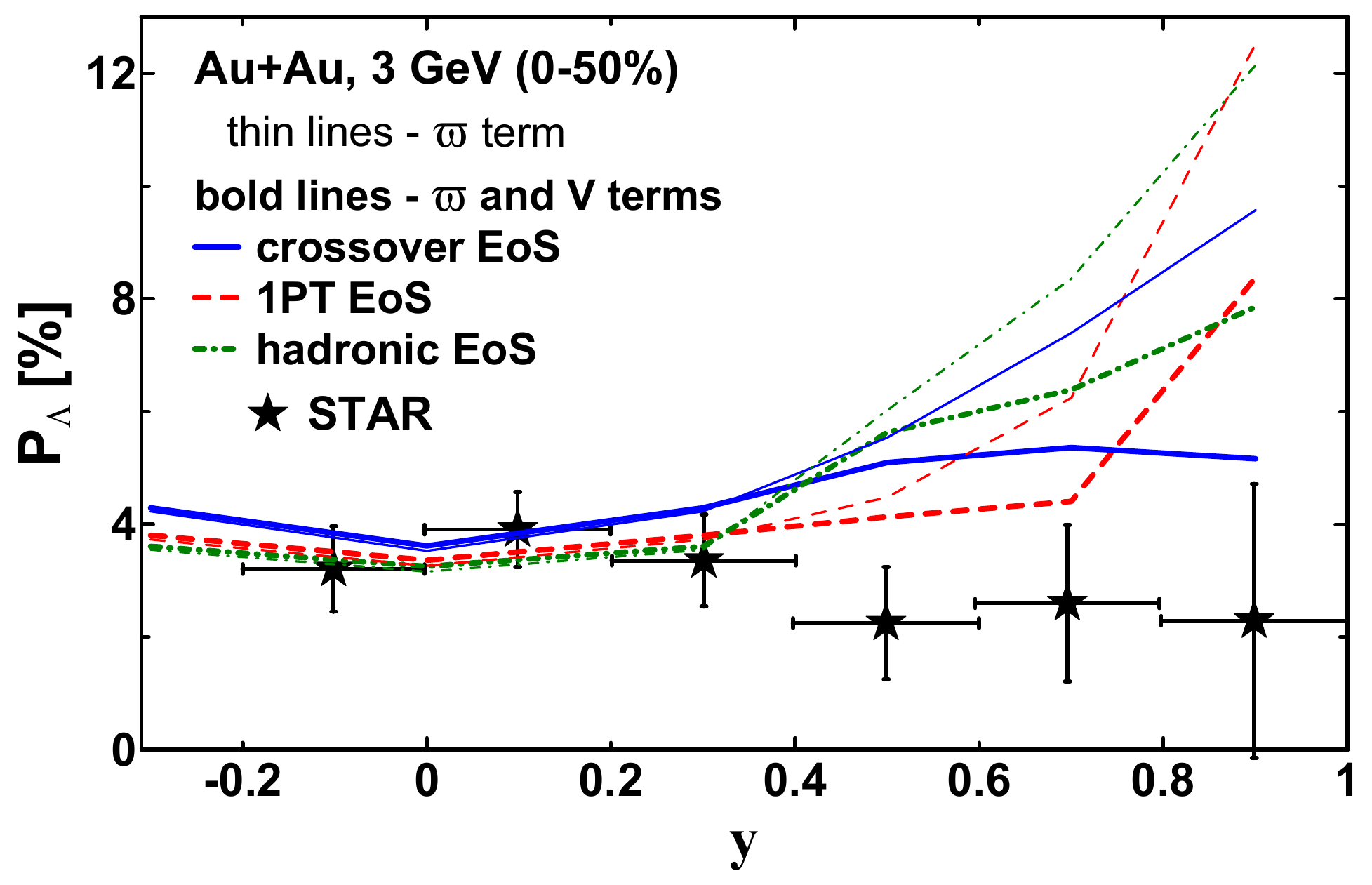}
 \caption{(Color online)
Rapidity dependence of the global $\Lambda$ polarization
in Au+Au collisions at  $\sqrt{s_{NN}}=$ 3 GeV (centrality 0-50\%),
originated from only the thermal vorticity (thin lines) 
and with additional vector-meson contribution (bold lines).
Results for different EoS's are presented.
 Data are from Ref. \cite{STAR:2021beb} (STAR). 
}
\label{fig6XV-3GeV}
\end{figure}

The rapidity dependence  of the global $\Lambda$ polarization
at  3 GeV, calculated accordingly to Eq. (\ref{PL-mean}),  is
shown in Fig. \ref{fig6XV-3GeV}. Both the thermal vorticity  
and with additional vector-meson contribution (bold lines in 
Fig. \ref{fig6XV-3GeV}) quite well describe the STAR data \cite{STAR:2021beb} 
at $|y|<0.3$. However, they overestimate the data at $|y|>0.3$.
The vector-meson contribution somewhat improves the agreement, 
especially with the crossover EoS, but the overestimation at $|y|>0.3$ 
persists. 

This observation demonstrates once again that effects of the thermal vorticity and 
vector-meson interaction   
become large in rapidity ranges overlapping with the
participant-spectator border, see Fig. \ref{fig3v}.
Moreover, the above contributions produce effects of opposite sign. 
Therefore, the observed  global $\Lambda$ polarization is 
a result of a delicate cancellation of the above contributions.

%
\begin{figure}[bht]
\includegraphics[width=8.5cm]{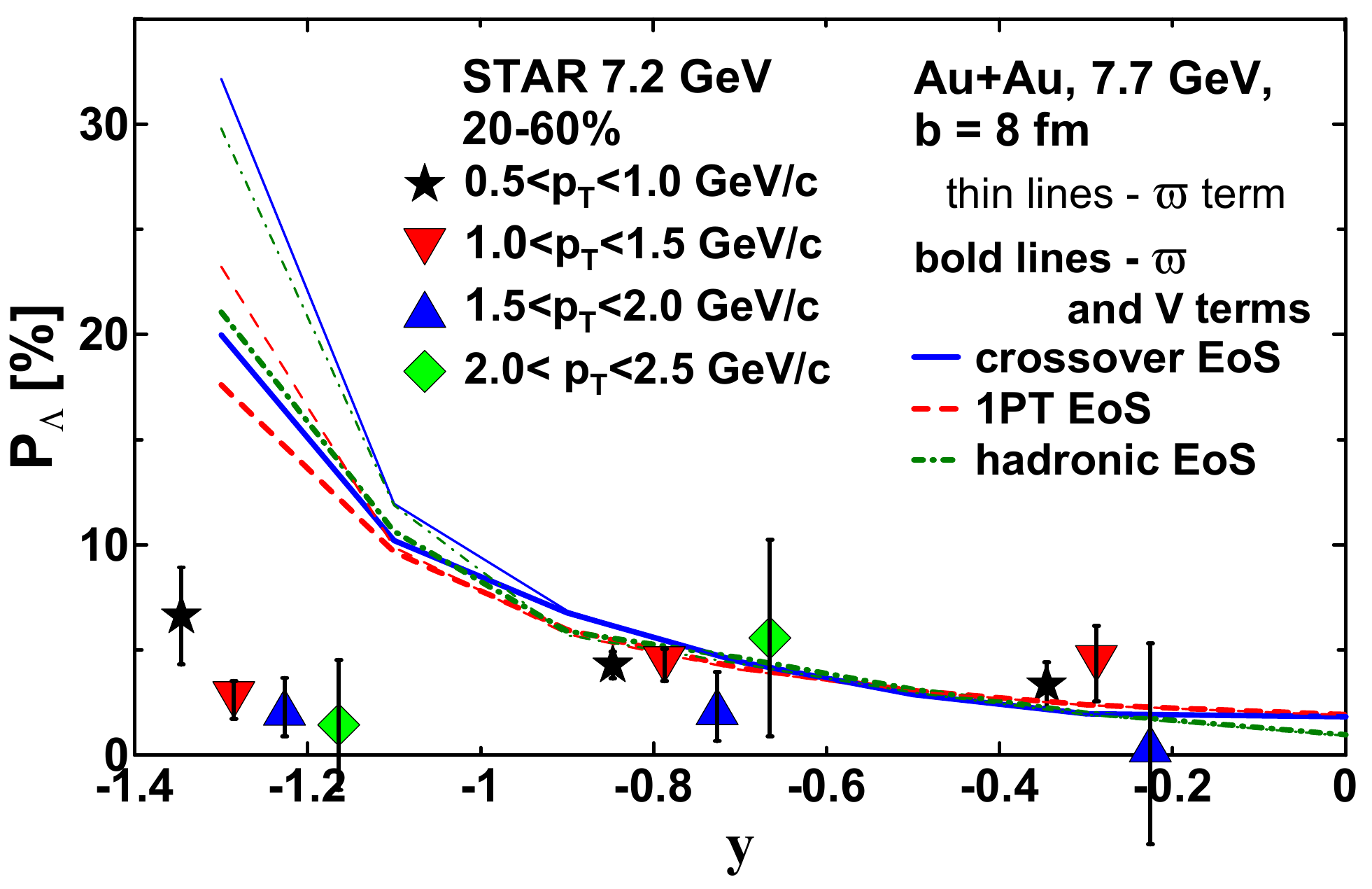}
 \caption{(Color online)
Rapidity dependence of the global $\Lambda$ polarization 
in Au+Au collisions at  $\sqrt{s_{NN}}=$ 7.7 GeV ($b=$ 8 fm), 
originated from only the thermal vorticity (thin lines) 
and that with additional vector-meson contribution (bold lines).
Results for different EoS's are presented.
Preliminary STAR data for Au+Au collisions at  $\sqrt{s_{NN}}=$ 7.2 GeV and centrality 20-60\%
are from Ref. \cite{Okubo:2021dbt}. 
}
\label{fig6XV-7GeV}
\end{figure}

The rapidity dependence  of the global $\Lambda$ polarization
at $\sqrt{s_{NN}}=$ 7.7 GeV is shown in Fig. \ref{fig6XV-7GeV}. It is compared 
with preliminary STAR data for Au+Au collisions at  $\sqrt{s_{NN}}=$ 7.2 GeV \cite{Okubo:2021dbt}. 
The STAR centrality selection is 20-60\%, which corresponds to the impact-parameter range 
$b=$ 5.6--9.7 fm  based on the overlap calculator \cite{web-docs.gsi.de}. Therefore, $b=$ 8 fm
can be chosen to represent this range. While the STAR data are presented by four subsets corresponding 
to different selections of transverse momentum, $p_T$, the 3FD results correspond to averaging over the 
whole $p_T$ range. 

In Fig. \ref{fig6XV-7GeV} we observe the same features as those for $\sqrt{s_{NN}}=$ 3 GeV 
in Fig. \ref{fig6XV-3GeV}. 
The thermal vorticity with or without additional vector-meson contribution 
well reproduces the STAR data \cite{Okubo:2021dbt} 
at $|y|<0.8$ but overestimates the data at $|y|>0.8$. The vector-meson contribution
somewhat reduces the disagreement with data at $|y|>0.8$.

%
\begin{figure}[bht]
\includegraphics[width=8.5cm]{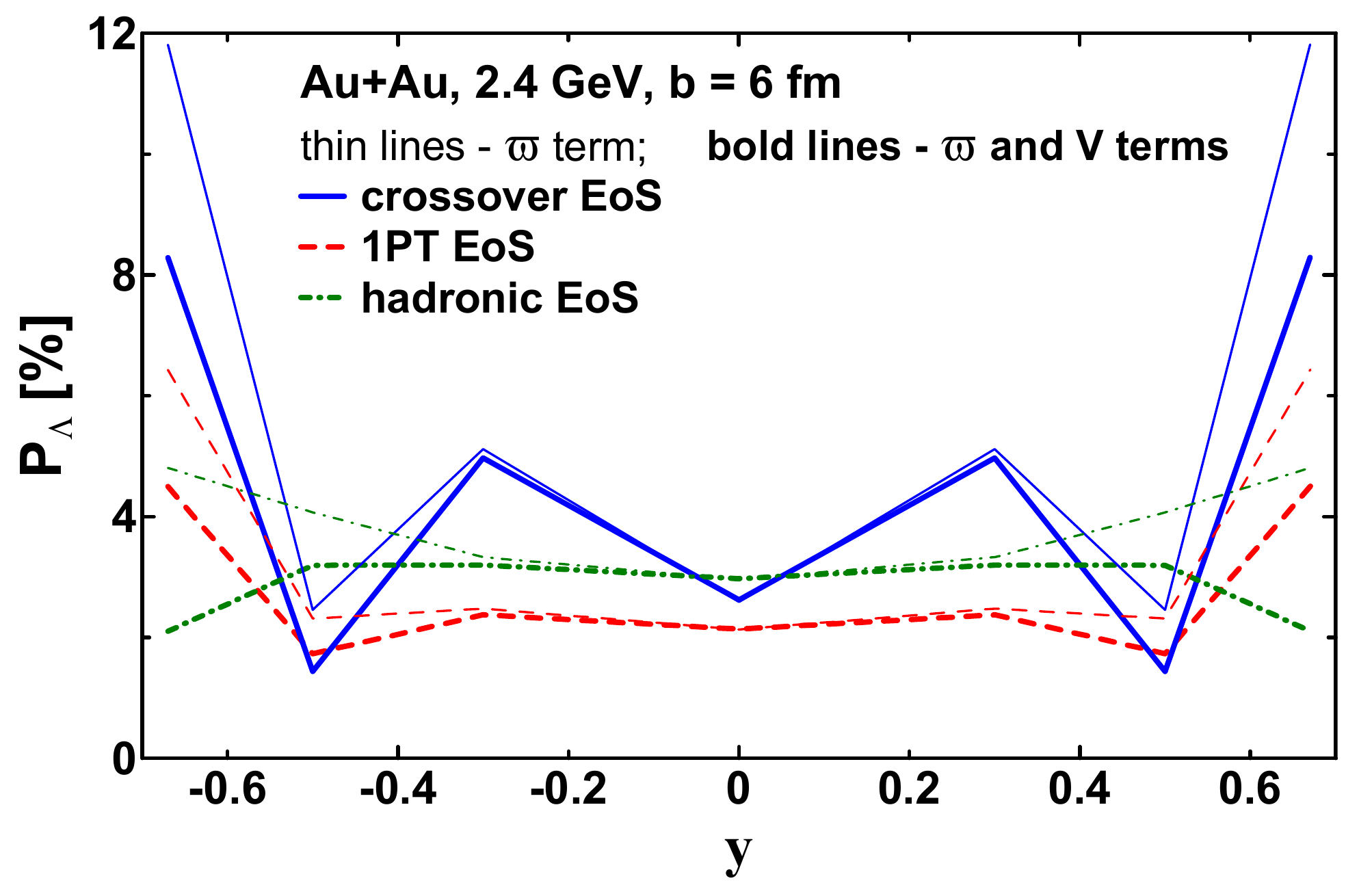}
 \caption{(Color online)
Rapidity dependence of the global $\Lambda$ polarization 
in Au+Au collisions at  $\sqrt{s_{NN}}=$ 2.4 GeV ($b=$ 6 fm), 
originated from only the thermal vorticity (thin lines) 
and with additional vector-meson contribution (bold lines).
Results for different EoS's are presented.
}
\label{fig6XV-2.4GeV}
\end{figure}

In Fig. \ref{fig6XV-2.4GeV} we present our predictions for the ongoing HADES 
experiment \cite{HADES:SQM2021}. We avoid modeling Ag+Ag collisions
because this system contains of too few particles, especially at low collision energies, 
to apply the hydrodynamical description. Therefore, we present predictions for the 
Au+Au collisions at $\sqrt{s_{NN}}=$ 2.4 GeV. The results for $b=$ 6 fm are shown, 
which approximately corresponds to centrality 10-40\%. As seen, the basic patterns
in Fig. \ref{fig6XV-2.4GeV} are the same as those in Figs. \ref{fig6XV-3GeV} and \ref{fig6XV-7GeV}. 
Only the difference of the results with different EoS's is larger.

\section{Centrality dependence}
\label{Centrality}

%
\begin{figure}[!pbht]
\includegraphics[width=8.5cm]{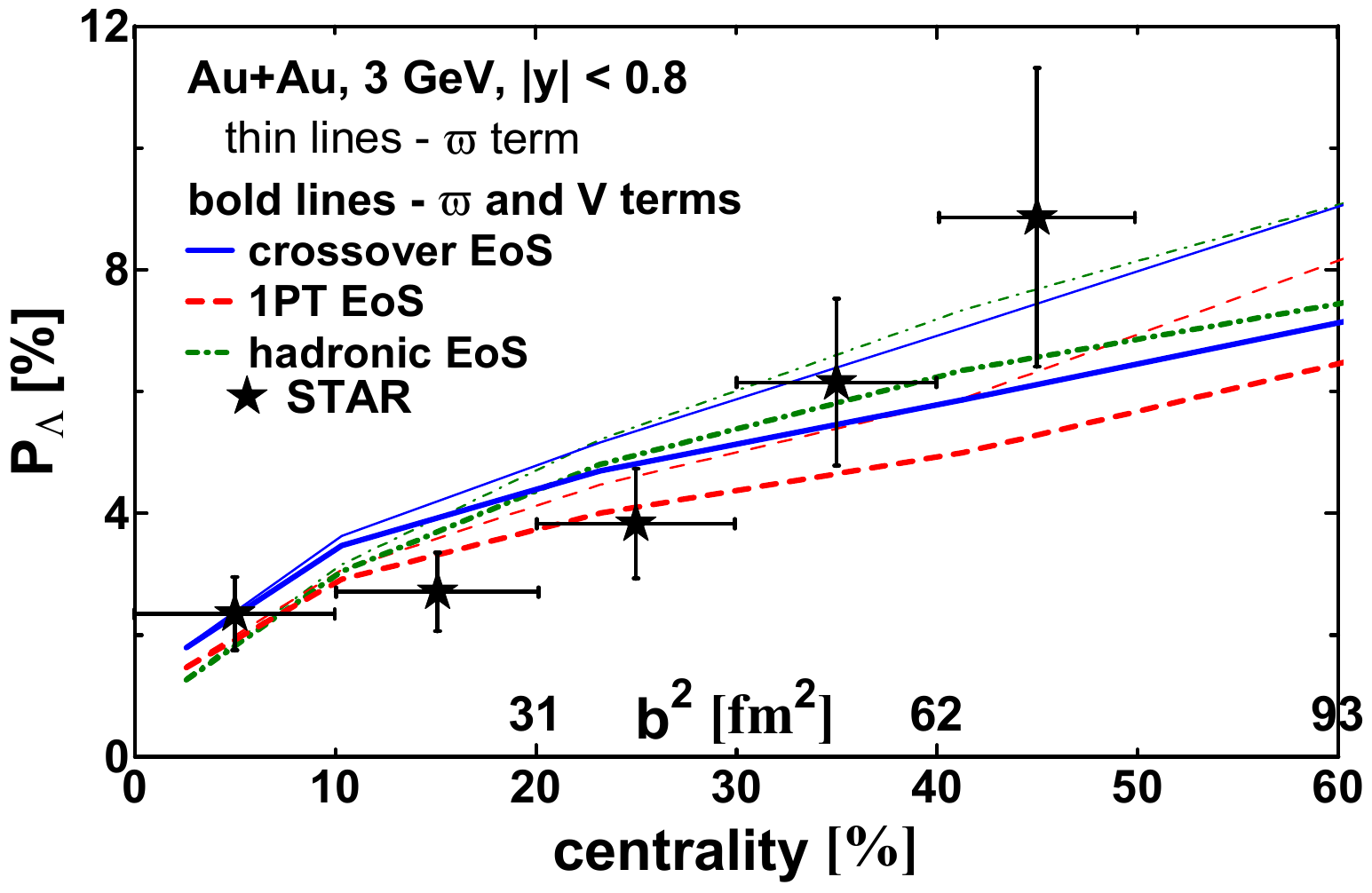}
 \caption{(Color online)
Centrality dependence of the global $\Lambda$ polarization in midrapidity region ($|y_h|<0.8$) 
in Au+Au collisions at  $\sqrt{s_{NN}}=$ 3 GeV  
originated from only the thermal vorticity (thin lines) 
and with additional vector-meson contribution (bold lines).
Results for different EoS's are presented.
 Data are from Ref. \cite{STAR:2021beb}. 
}
\label{fig7XV-3GeV}
\end{figure}
%
%
\begin{figure}[bht]
\includegraphics[width=8.5cm]{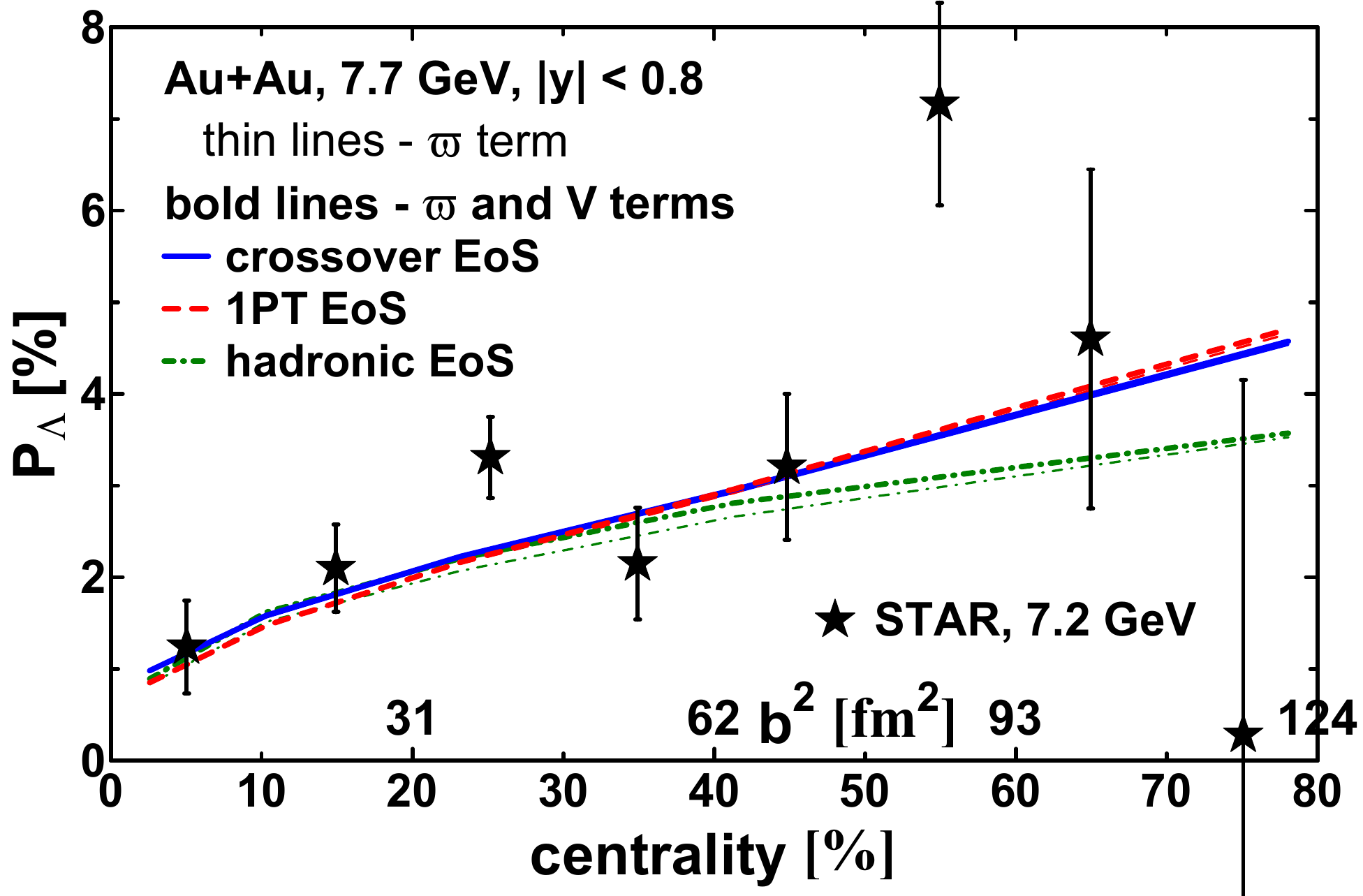}
 \caption{(Color online)
The same as in Fig. \ref{fig7XV-3GeV} but for $\sqrt{s_{NN}}=$ 7.7 GeV. 
Preliminary STAR data for Au+Au collisions at  $\sqrt{s_{NN}}=$ 7.2 GeV 
are from Ref. \cite{Okubo:2021dbt}, only statistical errors are displayed. 
}
\label{fig7XV-7GeV}
\end{figure}
%
%
\begin{figure}[bht]
\includegraphics[width=8.5cm]{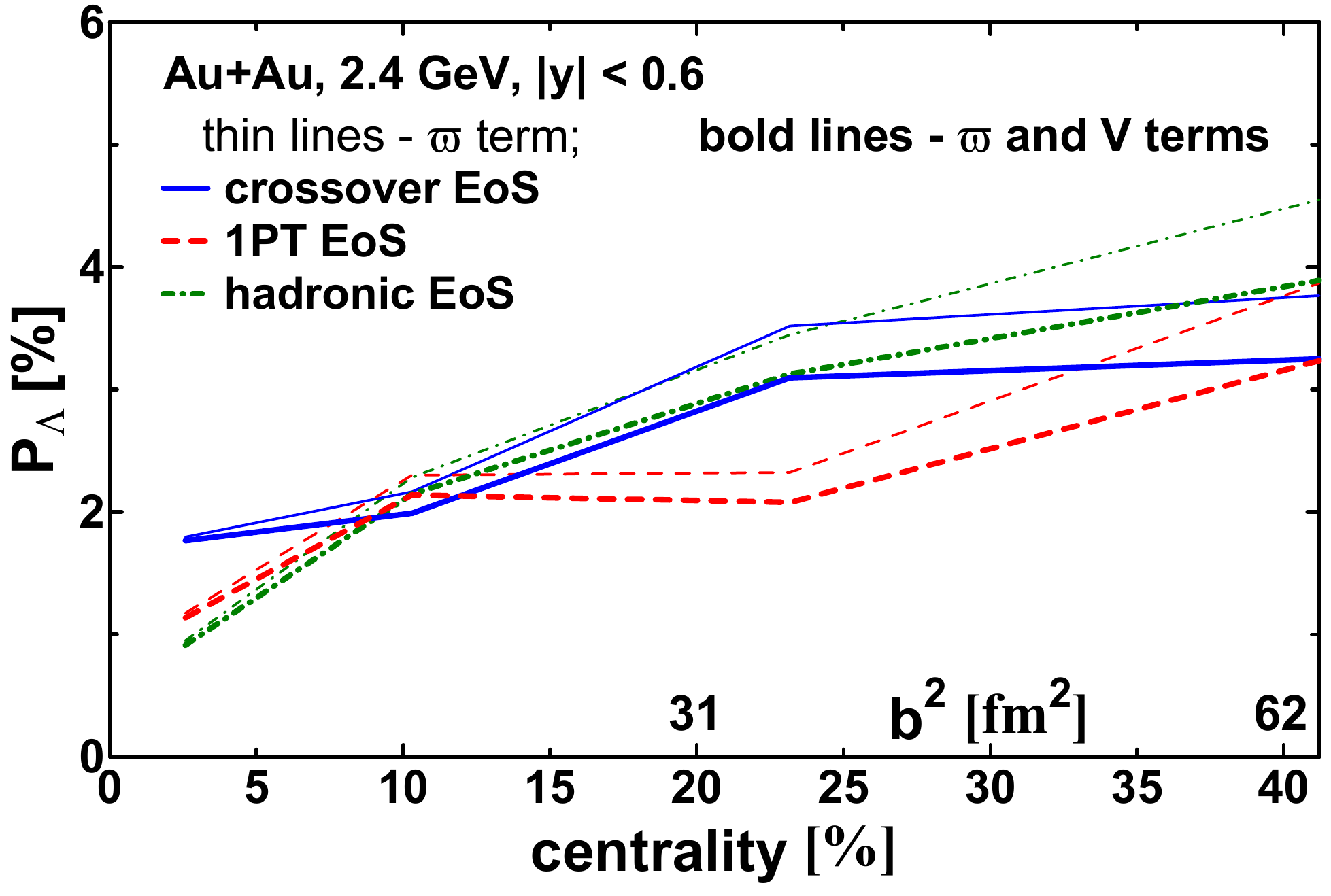}
 \caption{(Color online)
The same as in Fig. \ref{fig7XV-3GeV} but for the midrapidity region ($|y_h|<0.6$) 
at $\sqrt{s_{NN}}=$ 2.4 GeV. 
}
\label{fig7XV-2GeV}
\end{figure}
The global polarization of $\Lambda$ hyperons in Au+Au 
collisions  is calculated at impact parameters $b=$ 2, 4, 6, 8 and 11 fm.   
The displayed impact parameters  are associated with collision centrality by means of 
the Glauber simulations based on the nuclear overlap calculator \cite{web-docs.gsi.de}. 

In Fig. \ref{fig7XV-3GeV}, 
centrality dependence of the global $\Lambda$ polarization in midrapidity region ($|y_h|<0.8$) 
in Au+Au collisions at  $\sqrt{s_{NN}}=$ 3 GeV  
originated from only the thermal vorticity (thin lines) 
and with additional vector-meson contribution (bold lines) is displayed.
The experimental rapidity window is asymmetric -0.2 $<y<$ 1 \cite{STAR:2021beb}. 
However, it well complies with the modeled window in view of flat experimental rapidity dependence 
of the observed polarization, see Fig. \ref{fig6XV-3GeV}. 
As seen from Fig. \ref{fig7XV-3GeV}, 
both the thermal vorticity and that with additional vector-meson contribution 
reasonably well (though not perfectly) describe the observed centrality dependence.

At the energy of 7.7 GeV, see Fig. \ref{fig7XV-7GeV}, the effect of the additional vector-meson contribution
becomes negligible because the rapidity window does not cover the regions of the participant-spectator borders, 
as it has been already discussed in sect. \ref{Meson-field}. The thermal vorticity with and without the 
meson-field contribution reasonably well describes preliminary STAR data for 7.2 GeV
energy. 

Our predictions for the centrality dependence in Au+Au collisions 
at  $\sqrt{s_{NN}}=$ 2.4 GeV (HADES experiment) are presented in Fig. \ref{fig7XV-2GeV}. 
We took the rapidity window $|y_h|<0.6$, which is similar to that used for the 
Ag+Ag system in the  HADES experiment \cite{HADES:SQM2021}. 
Here the situation is similar to that at 3 GeV, see Fig. \ref{fig7XV-3GeV}, 
only the centrality dependence is weaker.

\section{EoS}
\label{EoS}

\begin{figure}[bht]
\includegraphics[width=7.5cm]{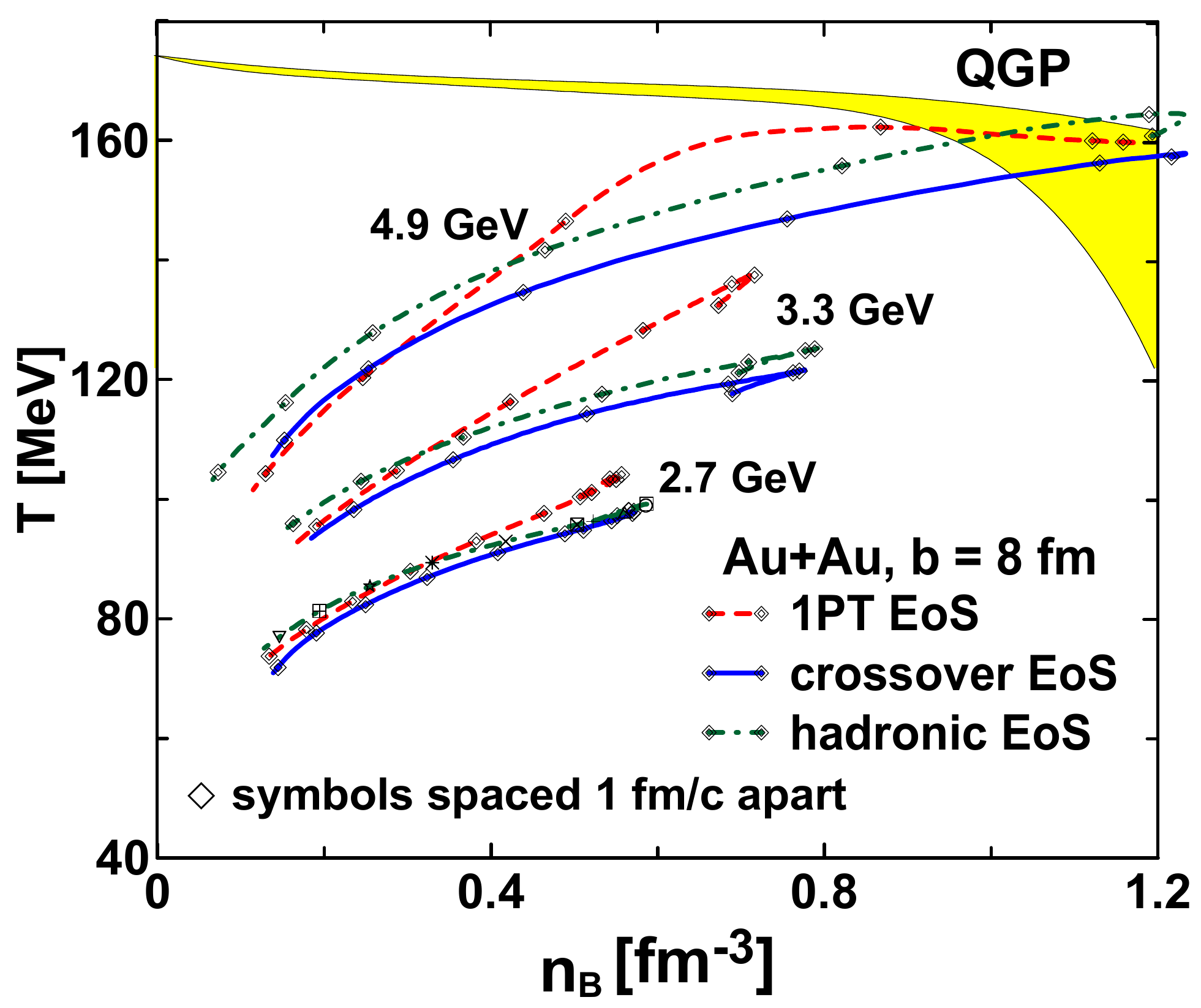}
 \caption{(Color online)
Dynamical trajectories of the matter
in the central box of the colliding nuclei (4fm$\times$4fm$\times$4fm$/\gamma_{cm}$),
where $\gamma_{cm}$ is the Lorentz factor associated with the initial
nuclear motion in the c.m. frame, for semi-central collisions ($b=$ 8 fm) of
Au+Au at $\sqrt{s_{NN}}=$  2.7, 3.3, 4.9 GeV. 
1PT, crossover and hadronic trajectories are displayed. 
The trajectories are plotted in terms of baryon density ($n_B$) and
temperature ($T$). Only expansion stages of the evolution are displayed. 
Symbols on the trajectories illustrate the expansion rate: they are spaced 1 fm/c apart. 
The yellow zone is a mixed-phase region within the 1PT scenario. 
}
\label{fig9}
\end{figure}

All the above presented calculations were performed with three EoS's. 
At moderately relativistic collision energies, $\sqrt{s_{NN}}\lsim$ 4.5 GeV, 
all these EoS's describe the hadronic matter, 
except for the crossover EoS containing the small QGP admixture even at low energies.  
This is seen from Fig. \ref{fig9}, where dynamical trajectories of the matter
in the central region of the colliding nuclei  in semi-central ($b=$ 8 fm) collisions of
Au+Au at $\sqrt{s_{NN}}=$  2.7, 3.3, 4.9 GeV are displayed. 
Only expansion stages of the evolution are displayed. 
The evolution proceeds from top-right to bottom-left.  
Symbols on the trajectories illustrate the expansion rate: they are spaced 1 fm/c apart. 
The yellow zone in Fig. \ref{fig9} is a mixed-phase region within the 1PT scenario. 
The critical temperature $T_c$ = 173 MeV for the 1PT EoS looks too high nowadays, 
cf. \cite{Borsanyi:2012cr}. This is because the 
1PT and crossover EoS's in Ref. \cite{Toneev06} were fitted to old, still imperfect lattice data 
\cite{Fodor:2002sd,Csikor:2004ik,Karsch:2000kv}. 
However, this shortcoming is not severe for the
reproduction of bulk observables in heavy-ion collisions.

%

In spite of that all the considered EoS's describe the hadronic matter
at moderately relativistic collision energies, they are not identical. 
Indeed, the corresponding dynamical trajectories in the hadronic phase  
are different, see Fig. \ref{fig9}, though close to each other. 
Therefore, differences in predictions of these EoS's can be considered 
as an uncertainty resulting from EoS ambiguity in the hadronic phase. 
All the considered EoS's give almost identical predictions for 
bulk \cite{Ivanov:2013wha,Ivanov:2013yqa,Ivanov:2013yla,Ivanov:2018vpw}
and even flow \cite{Ivanov:2014zqa,Konchakovski:2014gda,Ivanov:2014ioa}
observables at moderately relativistic collision energies. 
The polarization turns out to be more sensitive to details of the EoS.

\section{Summary}
\label{Summary}

Based on the 3FD model, the global $\Lambda$  polarization 
in Au+Au collisions at moderately relativistic energies, 
2.4 $\leq\sqrt{s_{NN}}\leq$ 7.7 GeV, was calculated, including its 
rapidity and centrality dependence.  
Contributions of the thermal vorticity and meson-field interaction \cite{Csernai:2018yok}
to the global  polarization were considered. 
Feed-down from higher-lying resonances was also studied, which as found  
reduces the polarization  by $\approx$25\% at lower energies and by $\approx$15\% at 7.7 GeV.
The results were compared with data from recent and ongoing experiments
\cite{STAR:2021beb,Okubo:2021dbt,HADES:SQM2021}.
It is predicted that 
the global polarization increases with the collision energy decrease. 
A maximum is reached at $\sqrt{s_{NN}}\approx$ 3 GeV, if the measurements are performed
with the same acceptance. 

The value of the polarization is very sensitive to interplay of the aforementioned different
contributions. In particular, the thermal vorticity predicts quite strong increase of the polarization
from the midrapidity to forward/backward rapidities, while 
the meson-field contribution considerably flattens the 
rapidity dependence. 
The meson-field contribution is large at the participant-spectator border and 
hence considerably reduces the polarization at forward/backward rapidities, while it 
is practically negligible at the midrapidity.  
As a rule, it improves agreement of calculated polarization with available data. 
Note that one of many possible parametrizations of the meson-field interaction was 
used in the present calculations. It indicates the order of magnitude and character of the produced 
effect. The details may be different for other, more refined parametrizations, e.g., such as those developed in 
Refs. \cite{Maslov:2015wba,Maslov:2015msa,Hornick:2018kfi} for astrophysical applications.

The simulations were performed with three different EoS's. 
In spite of that, all the considered EoS's describe the hadronic matter
at $\sqrt{s_{NN}}\lsim$  4.5 GeV, they are not identical. 
The polarization turns out to be more
sensitive to details of the EoS than bulk and even flow  observables.
The EoS crossover is somewhat preferable,  
although the data reproduction is far from being perfect. 
This could be a result of imperfectness of the crossover EoS, 
in view of high sensitivity of the global polarization to the EoS. 
Alternatively, this may indicate that the effect of the 
thermal-shear contribution \cite{Becattini:2021suc,Liu:2021uhn,Becattini:2021iol} 
should be additionally explored. 
Authors of Ref. \cite{Sun:2021nsg} found that at the energy of 19.6 GeV the effect of 
the thermal shear is negligibly small at the freeze-out stage.
Whether this is so at moderately relativistic energies remains to be seen.

All presently available approaches to the particle polarization, i.e. the thermodynamic approach 
used here  \cite{Becattini:2013fla,Becattini:2016gvu,Fang:2016vpj}
and that based on the chiral-vortical effect
\cite{Vilenkin:1980zv,Son:2004tq,Gao:2012ix,Sorin:2016smp,Baznat:2017jfj,Ivanov:2020qqe},  
require the thermal equilibrium at the freeze-out stage. 
At the same time the collision dynamics 
becomes less and less equilibrium with the collision energy decrease. This becomes a problem 
at low energies. We argue that the equilibrium is achieved at the freeze-out stage, only this 
equilibration takes longer.

\begin{acknowledgments} 
Helpful discussions with 
E.E.~Kolomeitsev and D.N. Voskresensky are gratefully acknowledged. 
This work was carried out using computing resources of the federal collective usage center ``Complex for simulation and data processing for mega-science facilities'' at NRC "Kurchatov Institute", http://ckp.nrcki.ru/.
Y.B.I. was partially supported by the Russian Foundation for
Basic Research, Grants No. 18-02-40084 and No. 18-02-40085. 
This work was also supported by MEPhI within the Federal Program "Priority-2030".

\end{acknowledgments}

\end{document}